\documentclass[]{aa}  

\usepackage{graphicx}
\usepackage{subfigure}
\usepackage{adjustbox}
\usepackage{txfonts}
\begin{document}

   \title{A case study of ACV variables discovered in the Zwicky Transient Facility survey}

   \author{N. Faltov{\'a}\inst{1}
   \and K.~Kallov{\'a}\inst{1}
   \and M.~Pri{\v s}egen\inst{1}
   \and P.~Stan{\v e}k\inst{1}
   \and J.~Sup{\'i}kov{\'a}\inst{2}
   \and C.~Xia\inst{1}
   \and K.~Bernhard\inst{3,4} 
   \and S.~H{\"u}mmerich\inst{3,4}
   \and E.~Paunzen\inst{1}}

   \institute{Department of Theoretical Physics and Astrophysics, Faculty of Science, Masaryk University, Kotl\'{a}\v{r}sk\'{a} 2, 611 37 Brno, Czech Republic
    \and Faculty of Informatics, Masaryk University, Brno, Czech Republic 
    \and Bundesdeutsche Arbeitsgemeinschaft f{\"u}r Ver{\"a}nderliche Sterne e.V. (BAV), Berlin, Germany
    \and American Association of Variable Star Observers (AAVSO), Cambridge, USA}

   \date{}

\date{}
 
  \abstract
   {Magnetic chemically peculiar (mCP) stars exhibit complex atmospheres that allow the investigation of the interplay of atomic diffusion, magnetic fields, and stellar rotation. A non-uniform surface distribution of chemical elements and the non-alignment of the rotational and magnetic axes result in the variability of several observables. Photometrically variable mCP stars are referred to as $\alpha^2$ Canum Venaticorum (ACV) variables.}
   {The present work presents a case study of known variables from the Zwicky Transient Facility (ZTF) survey, with the aim of investigating the survey's suitability for the detection and study of new ACV variables.}
   {Using suitable selection criteria based on the known characteristics of ACV variables, candidate ACV stars were selected from the ZTF Catalog of Periodic Variable Stars. All light curves were inspected in detail to select the most promising candidates. Where available, low-resolution spectra from the Large Sky Area Multi-Object Fiber Spectroscopic Telescope (LAMOST) were employed to classify the stars on the MK system and confirm their status as mCP stars.}
   {We have identified 86 new promising ACV star candidates. 15 of these stars have LAMOST spectra available, which, in all cases, confirm them as classical mCP stars, which highlights the viability of our approach. We present astrophysical parameters for all sample stars, which can be sorted into four subgroups characterized by distinct light curve shapes. Anti-phase variations in different photometric passbands, in particular, is a unique characteristic of a subset of ACV stars readily usable for their identification. The availability of data in three different passbands ($g$, $r$, and $i$) is a major advantage of the ZTF survey.}
   {On the basis of our experience with other photometric surveys and the analysis of light curves, we conclude that the ZTF is well suited for the search for, and the analysis of, ACV variables, which, however, are not considered in the available ZTF variable star catalogues. Further work will be concerned with the development and refinement of a search algorithm to correctly identify these stars in ZTF data and, subsequently, in massive photometric time-series databases in general.}


   \maketitle
%

\section{Introduction} \label{introduction} 

The magnetic chemically peculiar (mCP) stars, which encompass the groups of the Ap/CP2 stars and the He-peculiar stars \citep{preston_1974}, are upper-main sequence objects (spectral types B to early F) that exhibit anomalous surface abundances 
and a non-uniform distribution of chemical elements. They are furthermore characterized by the presence of a strong and global magnetic field. The CP2 stars show overabundances of Si, Sr, Cr, Eu, and the rare-earth elements as compared to the solar composition \citep{preston_1974,saffe2005}. The He-peculiar stars comprise of the B5 to B9 He-weak (CP4) stars that show anomalously weak \ion{He}{i} lines for their temperature type and the more massive B1 to B3 He-strong stars with anomalously strong \ion{He}{i} lines \citep{1998A&A...337..512A}.

In addition, there are also the CP1 stars 
\citep[the metallic-line or Am/Fm stars;][]{2019ApJS..242...13Q} 
and the CP3 stars \citep[the mercury-manganese or HgMn stars;][]{2021A&A...645A..34P}.
These groups are believed to possess only very weak, if any, magnetic fields \citep{2003A&A...407..631B}.

The non-uniform distribution of chemical elements characteristic for mCP stars is responsible for 
the formation of spots and patches of enhanced element abundances on their surfaces \citep{michaud1981}. Several mechanisms and their interplay may be responsible for the development of these spots, such as a weak stellar wind \citep{krticka2016}, the characteristics of the stellar magnetic field \citep{romanyuk2014}, the
occurrence of a weak convective zone, and the implications of slow rotation.

As a result of the abundance inhomogeneities, the rotation of mCP stars necessarily results in periodic variations of the observed flux ('photometric' spots; \citealt{krticka2007}) and periodic changes in their spectra. This variability can be explained by the oblique rotator model, which was proposed by \citet{babcock1949} and developed by \citet{Stibbs1950}. The period of the observed variations is hence identical with the rotational period of an mCP star. $\alpha^2$ Canum Venaticorum was the first CP2 star that was identified as a photometric variable \citep{GuthnikPrager1914}, which is the reason why photometrically variable mCP stars are generally referred to as $\alpha^2$ Canum Venaticorum (ACV) variable stars \citep{samus}.

Several elements play a part in the observed variations. For instance, \citet{krticka2007} demonstrated that the variability of the He-strong star HD 37776 is caused by spots of He and Si, whereas \citet{krticka2009} produced evidence that the variability of the CP2 star HR 7224 is the result of an inhomogeneous surface distribution of Fe and Si. Other elements such as Si, Cr, Sr, or Eu may contribute to the observed variations in CP2 stars \citep{bernhard2015b}.

ACV variables are perfectly suited for a direct measurement of the rotational period without the need for any additional calibrations. With astrophysical parameters such as stellar masses, effective temperature, radii, inclinations, and critical rotational velocities, it is possible to analyse the conservation of angular momentum during the main-sequence evolution
\citep{2017MNRAS.468.2745N}.

Although many excellent data sources on stellar variability are now available, the class of the ACV variables is only rarely considered in the algorithms employed for the automatic classification of variable stars into astrophysically meaningful classes \citep{2010ApJ...713L.204B,2015MNRAS.451.3385K,2016MNRAS.459.3721B}. In this paper, we present a case study of new ACV variables discovered in data from the Zwicky Transient Facility (ZTF) and show that the accuracy and cadence of the ZTF allows to detect previously unknown stars of this kind. This survey reaches to fainter magnitudes than most corresponding programs \citep{2018CoSka..48..277H}, which means that we can probe the population of ACV variables at larger distances from the Sun. Up to now, the automatic detection and classification methods employed with the ZTF data do not include this class of variable stars \citep{2020ApJS..249...18C,2020MNRAS.499.5782O}.

   \begin{figure}
   \centering
   \includegraphics[width=\hsize]{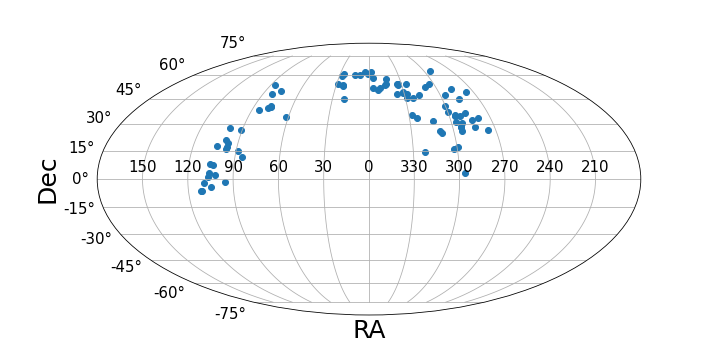}
   \includegraphics[width=\hsize]{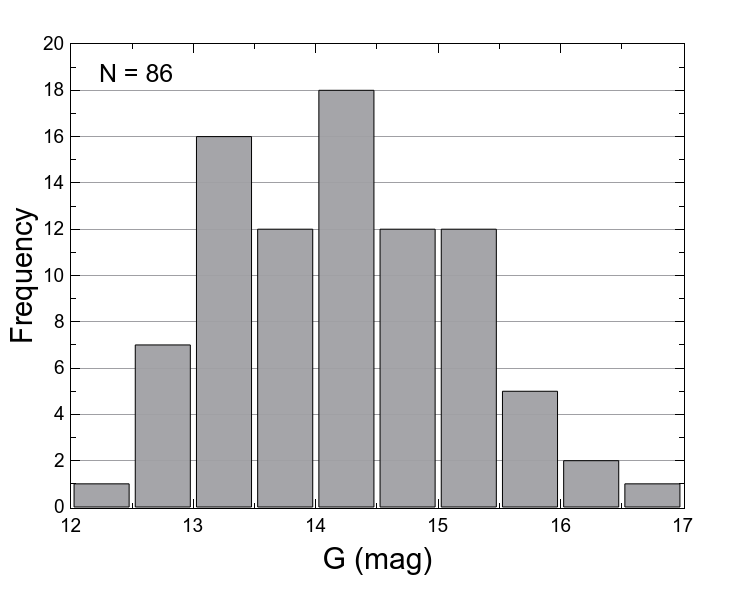}
      \caption{Mollweide projection in right ascension and declination (upper panel) and the $G$ magnitude distribution (lower panel) of our sample stars. All stars are located within the Galactic disk.}
         \label{star_sample_moll}
   \end{figure}
   
  \begin{figure*}
   \centering
   \includegraphics[width=\hsize]{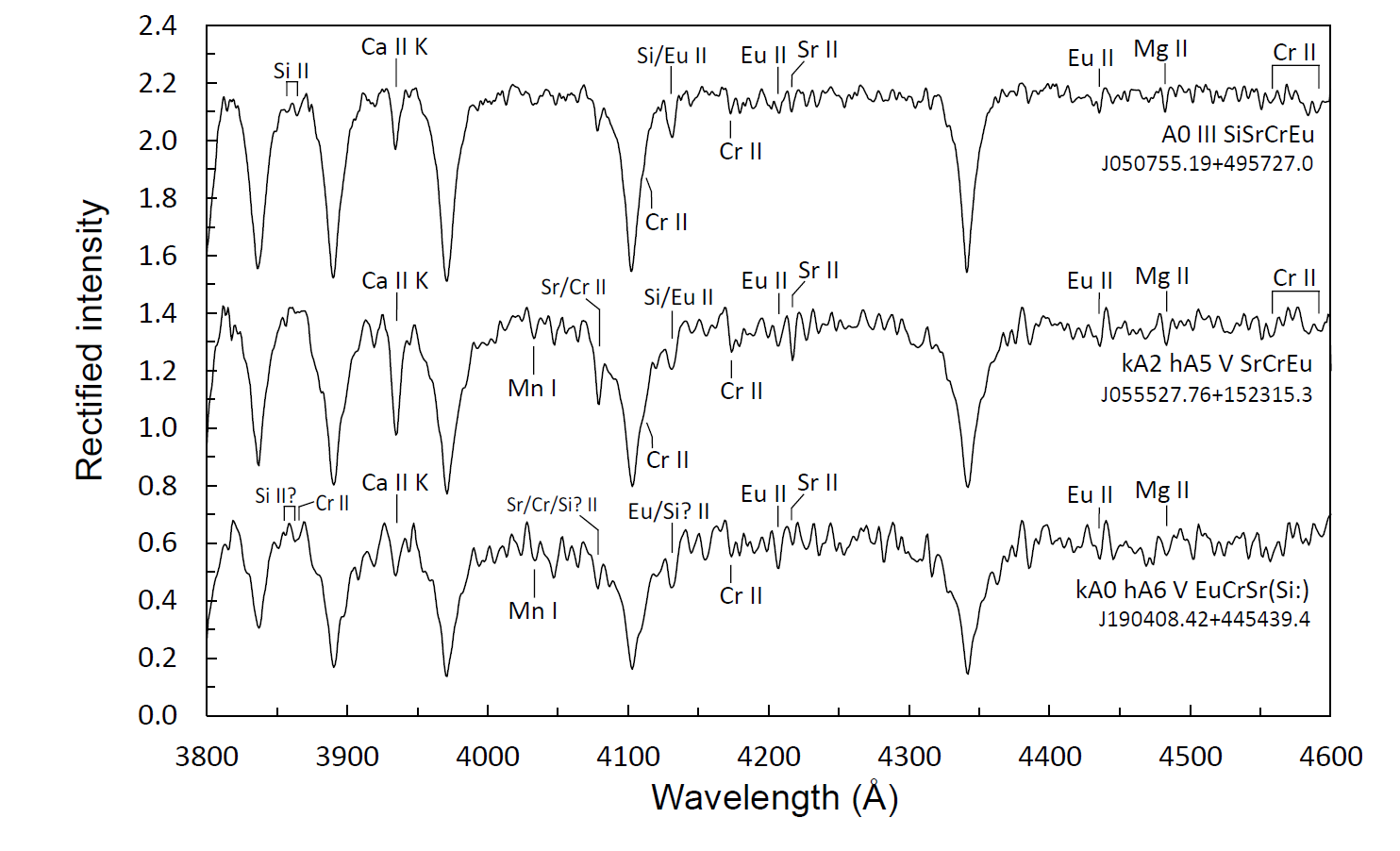}
      \caption{Blue-violet spectra region of three newly identified mCP stars, based on spectra from LAMOST DR7. Some prominent lines of interest are identified.} 
         \label{showcase}
   \end{figure*}

\section{Data selection and time series analysis} \label{data_selection}

The ZTF is a time-domain survey that is located 
at Palomar Observatory and in operation since 2017. It is a successor of the successful Palomar Transient Factory (PTF). The ZTF camera employs e2v CCD231-C6 devices and is mounted on the Palomar 48-inch Samuel Oschin Schmidt Telescope. It is scanning 3750 square degrees an hour in three different passbands ($g$, $r$, and $i$) to a limiting magnitude of 20.5 mag. The main aim is to discover young supernovae as well as
other types of transients. With its current strategy, the ZTF survey provides nearly 300 observations each year per object in the northern hemisphere. ZTF data are therefore ideally suited for studies of variable stars, binaries, AGN, and asteroids \citep{2019PASP..131a8002B,masci_2019}.

\citet{2020ApJS..249...18C} exploited the ZTF Data Release 2 and compiled a catalogue of 781\,602 periodic variable stars (The Zwicky Transient Facility Catalog of Periodic Variable Stars\footnote{\url{http://variables.cn:88}}).We systematically searched this catalogue for candidate ACV variables by examining the light curves of an
extended sample of stars in detail. Since the catalog does not contain a category for ACV-type variables, it can be assumed that any such objects were assigned to other categories of short-period variables, especially to the class of the RS Canum Venaticorum (RS CVn) stars, on which we concentrated in our analysis. RS CVn stars are rotational variables whose light curves, at least on first impression, look similar to the light curves of ACV variables \citep{2015ARep...59..937K}. They are, however, close binary stars with active chromospheres and enhanced spot activity \citep{1993AJ....106.1181E}.

As there are 81\,393 RS CVn stars listed in \citet{2020ApJS..249...18C}, further restrictions were made based on the known characteristics of the group of ACV variables \citep{2017MNRAS.468.2745N,2019A&A...622A.199J}:

\begin{itemize}
    \item[(i)] photometric period between one and ten days,
    \item[(ii)]  amplitude in the $r$ band smaller than 0.3\,mag,
    \item[(iii)]  the presence of a single independent variability frequency and corresponding harmonics,
    \item[(iv)]  stable or marginally changing light curve throughout the covered time span,
    \item[(v)]  $T_{\rm eff}$ \citep[taken from][]{2018A&A...616A...8A} between 6000\,K and 25\,000\,K.
\end{itemize}

Items (iii) and (iv) are helpful to distinguish ACV variables from, respectively, pulsators (which are often multi-periodic) and other types of rotational variables, such as RS CVn stars. In contrast to ACV variables, whose light curves remain stable over very long-periods of time (decades or more), most spotted stars show significant light changes as spots form and decay.

1400 objects passed these selection criteria, whose light curves were downloaded from the ZTF website\footnote{\url{https://www.ztf.caltech.edu/}} and visually inspected. The most promising ACV candidates were selected, which resulted in the 86 candidates presented in this paper. The big advantage of the ZTF survey is that photometric data are available in at least two filters, which facilitates the differentiation from other similar groups of variables stars, such as the ellipsoidal variables.

The published periods by \citet{2020ApJS..249...18C}
were inspected using the program package {\sc Period04} \citep{2005CoAst.146...53L}, which performs a discrete Fourier transformation. Furthermore, we used 
the {\sc cleanest} and phase dispersion minimization (PDM) algorithms as implemented in the program package {\sc Peranso} \citep{2016AN....337..239P}. The same results were obtained within the derived errors, which depend on the time series characteristics, i.e. the distribution of the measurements over time and photon noise.

The complete set of phased light curves in the available filters is presented in the Appendix (Table \ref{light_curves_all}).
   
\begin{table}
\begin{center}
\caption{Spectral types of the 15 sample stars with spectra in LAMOST DR7. Also indicated are the LAMOST observation ID (Obs\_ID) and the signal-to-noise ratio in the $g$ band (SN).}
\label{result_classification}
\tiny
\begin{tabular}{cccl}
  \hline
LAMOST\_ID & Obs\_ID & SN & spectral type \\
\hline
J012145.51+443625.8	&	168003189	&	98	&	kA1 hA5 V SrEu(Cr)	\\
J040517.81+335007.7	&	189415040	&	37	&	kA0 hA5 V EuSrCr	\\
J050755.19+495727.1	&	290903161	&	64	&	A0 III SiSrCrEu	\\
J051331.05+393032.4	&	294302231	&	103	&	kA1 hA7 V SrEuCr	\\
J052856.69+475711.7	&	253703006	&	67	&	kA1 hA7 V SrCrEu	\\
J053725.00+382012.2	&	316409236	&	87	&	A0 III SiEuSrCr	\\
J054009.21+120945.0	&	476111206	&	106	&	kA2 hA3 V SrCrEu	\\
J055527.76+152315.3	&	98805041	&	88	&	kA2 hA5 V SrCrEu	\\
J065551.61+174248.1	&	372816239	&	75	&	kA1 hA7 V SrCrEu	\\
J182441.87+485404.6	&	155808159	&	17	&	kA0: hA5: V SrCrEu	\\
J190408.38+445439.1	&	362115243	&	110	&	kA0 hA6 V EuCrSr(Si:)	\\
J190831.72+511959.7	&	575716189	&	153	&	kA0 hA3 V EuSrCr(Si:)	\\
J194356.39+470831.0	&	572806020	&	96	&	kA1 hA5 V SrEu(Cr)	\\
J212905.49+142755.4	&	592409127	&	113	&	kA1 hA7 V SrCrEu(Si:)	\\
J213347.82+451647.6	&	476310217	&	51	&	kA2 hA7 V SrCrEu	\\
\hline
\end{tabular}
\end{center}
\end{table}
   
\begin{figure*}
   \centering
   \includegraphics[width=\hsize]{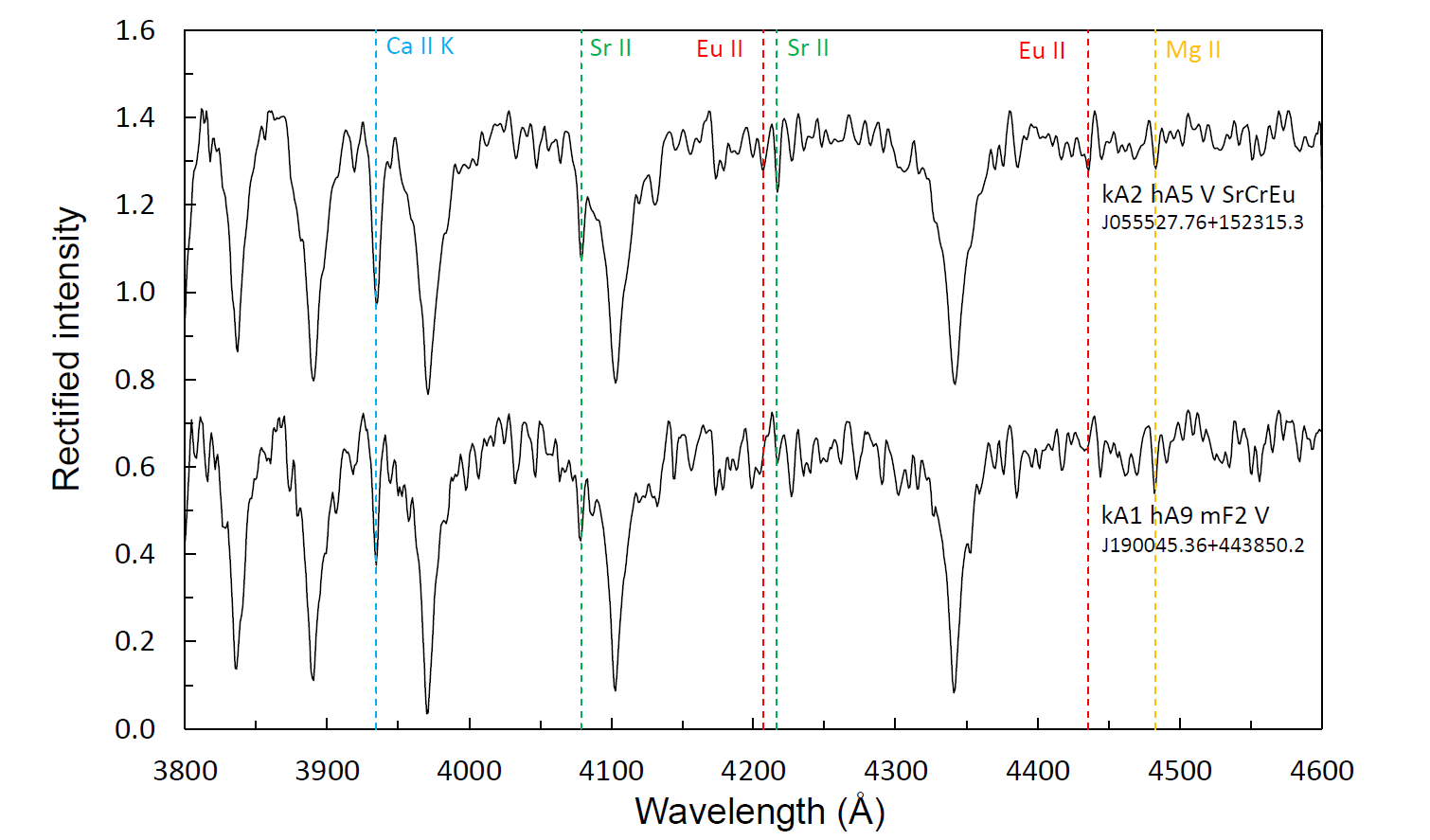}
      \caption{The spectra of a CP2 star (upper panel) of the present sample and a CP1 star (lower panel) taken from \citet{2019ApJS..242...13Q} in the traditional blue-violet spectral region. Some prominent lines of interest are identified.} 
         \label{Am_Ap}
   \end{figure*}

\section{Sample characteristics} \label{sample}

In the upper panel of Fig. \ref{star_sample_moll}, the distribution of the stars on the sky is shown. All stars are located in the Galactic disk with distances between 1000 and 6000 pc from the Sun. They are also, on average, fainter (Fig. \ref{star_sample_moll}, lower panel) than the known ACV variables investigated with other ground-based surveys \citep{bernhard2015a,bernhard2015b,2020MNRAS.493.3293B}.

\subsection{Spectral classification} \label{spectral_classification}

In order to validate the selection process of our target star sample (Section \ref{data_selection}) and assess its efficiency, we searched for matches in DR7 of the Large Sky Area Multi-Object Fiber Spectroscopic Telescope (LAMOST) database \citep{lamost}.

The LAMOST project employs a 4 meter quasi-meridian 
reflecting Schmidt telescope capable to observe up to 4000 targets per exposure simultaneously in 
a field of view of 5$^{\circ}$ \citep{cui_2012}. Recently, in March 2020, the DR7 released more than 10 million low-resolution spectra in the wavelength range from 3690$-$9100 $\AA$, with a resolution of R\,$\sim$\,1800.

In total, we found spectra for 15 stars from which five
(ZTF J040517.81+335007.7, J050755.19+495727.2, J054009.21+120945.1, J183551.46+334749.7,
and J212905.49+142755.5) have more than one spectrum available. Only the spectrum with the highest signal-to-noise ratio was considered for the analysis in these cases. We have to emphasize that the available
sample is not biased by the spectral type range but by the apparent magnitude. It should therefore be a representative sample of 86 ACV stars.

The spectral classification was performed in the framework of a revised MK system by visually comparing, with the help of a simple graphical program, spectra in the spectral range from 3700\,\AA\ to 9100\,\AA\ to the spectra of standard stars using the notation of \citet{1987ApJS...65..581G,1989ApJS...69..301G,1989ApJS...70..623G,1994AJ....107.1556G}. As ACV stars are mainly early B- to early F-type main sequence stars, the spectral range of the standard star spectra was chosen to extend from 3800\,\AA\ to 4600\,\AA, because the majority of the relevant spectral lines is situated in this wavelength region.

In the spectral classification, we focused on several relevant indicators: the profile of the H lines, the ratio between the \ion{Ca}{ii} K and the H$\varepsilon$ line, and the presence and strength of several well known metallic lines \citep{2020A&A...640A..40H}. The spectra of all of our sample stars shows the characteristic signature of mCP stars -- the well-known flux depression at around 5200\,\AA\ \citep{2003MNRAS.341..849K,2004MNRAS.352..863K}. The \ion{Ca}{ii} K/H$\varepsilon$ line ratio lies in the range of 0.3 -- 0.75 for all of our sample stars, which is typical for the spectral classes A0 to A5. In most cases, the H lines present the broad profiles of luminosity class V objects, confirming that the majority of our sample stars belong to the main sequence. The two hotter stars of our sample (J050755.19+495727.1 and J053725.00+382012.2) show narrow H-line profiles best fit by standards of luminosity class III.

The spectral classes of our sample are listed in Table \ref{result_classification}. Peculiarities are listed in order of importance. Peculiarity types in brackets or followed by a colon denote, respectively, mild or questionable peculiarities. With the two exceptions mentioned above, all objects belong to the cool mCP stars, which is also reflected by their astrophysical parameters (Sect. \ref{astrophysical_parameters}). Example spectra are provided in Fig. \ref{showcase}.

Cool CP2 stars and CP1 stars overlap in the temperature domain, and care needs to be taken in the 
differentiation of their spectra, which may look similar on first glance. CP1 stars are characterized by 
significant underabundances of Ca and Sc and general overabundances of iron-peak and heavier elements 
\citep{2009ssc..book.....G}. Therefore, in relation to the spectral type as derived from the hydrogen lines, 
they show weak \ion{Ca}{ii} K lines and a pronounced metallic-line spectrum, which also applies to 
most cool CP2 stars.

Whereas the enhancement of heavy elements in CP1 stars is general, selected elements generally dominate the 
spectra of CP2 stars. Strong lines of the iron-peak element Cr are a common feature of both groups; strongly 
enhanced lines of, for example, Si and Eu, however, are a diagnostic feature of CP2 stars, which also generally 
show broader and more diffuse \ion{Ca}{ii} K lines. Figure \ref{Am_Ap} illustrates the subtle differences between the spectra 
of a CP1 star \citep{2019ApJS..242...13Q} and a CP2 star of the present sample in the traditional 
blue-violet spectral 
region. While \ion{Sr}{ii} 4077\,\AA\ is strong in both stars, the \ion{Sr}{ii} 4126\,\AA\ line, which is used to confirm 
Sr peculiarity \citep{2009ssc..book.....G}, is only enhanced in the CP2 star.

Most important, however, is the morphology of the 5200\,\AA\ region. CP2 stars show a characteristic flux 
depression around 5200\,\AA\, which is not observed in CP1 stars \citep{2005A&A...441..631P} and therefore an 
important criterion to 
differentiate these groups of objects.

Some of the stars, for example LAMOST J190408.37+445439.3, are very peculiar objects whose general spectral pattern resembles that of HD 101065 (Przybylski's Star), which is a rapidly oscillating Ap (roAp) star with a unique abundance pattern showing, for example, lines of \ion{Pm}{i}, \ion{Pm}{ii}, \ion{Tc}{i}, and \ion{Tc}{ii} 
\citep{2018MNRAS.477.3791H}. The cadence and accuracy of the ZTF measurements does not allow to search for roAp pulsations, which have periods below 30 minutes and amplitudes of only a few mmags \citep{2019MNRAS.487.3523C}. Unfortunately, most of our sample stars are also too faint to be observed by the TESS satellite \citep{2019MNRAS.487.4695S}. Therefore, new ground-based observations are needed to search for this type of variability in our sample stars \citep{2015A&A...575A..24P,2018RAA....18..135P}.

   \begin{figure}
   \centering
   \includegraphics[width=\hsize]{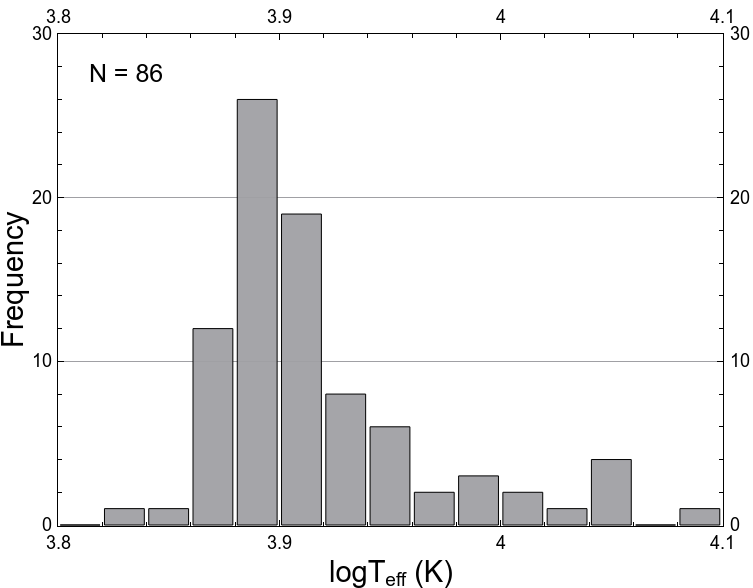}
   \includegraphics[width=\hsize]{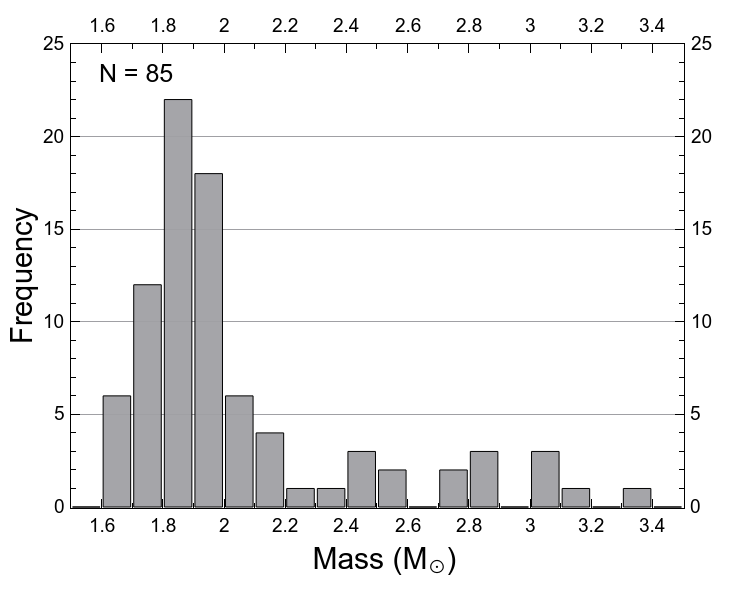}
   \includegraphics[width=\hsize]{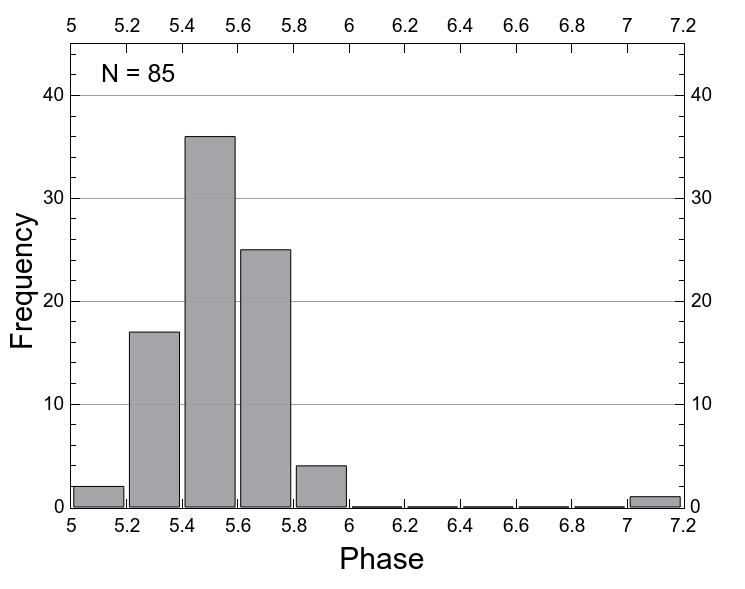}
      \caption{Distribution of determined effective temperatures (upper panel), estimated masses (middle panel), and estimated evolutionary phases (lower panel). The labelling in the lower panel follows the format of the employed models \citep{2012MNRAS.427..127B}, in which phase 5 represents the zero-age main sequence (ZAMS) and phase 6 the terminal-age main sequence (TAMS). Note that the large uncertainty in luminosity for one star (ZTFJ232952.58+501457.3) prevented us from calibrating corresponding mass and phase values.}
         \label{histograms}
   \end{figure}
   
   \begin{figure}
   \centering
   \includegraphics[width=\hsize]{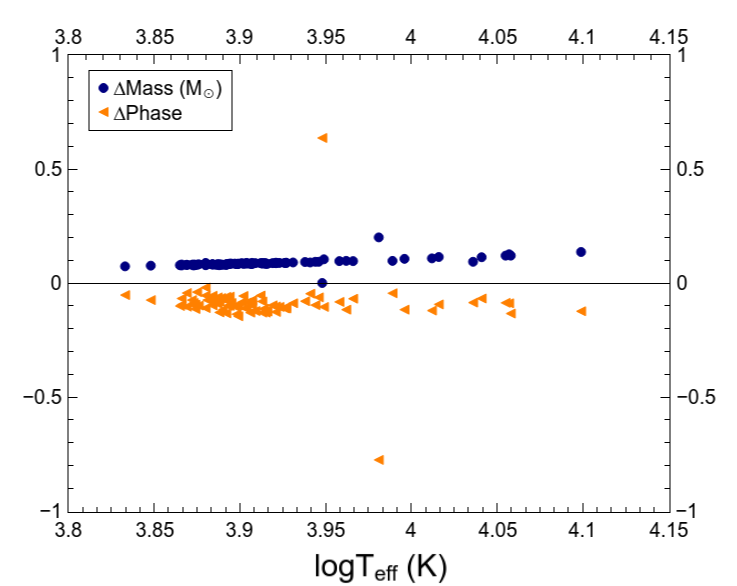}
      \caption{Comparison of estimated masses and evolutionary phases for [Z]\,=\,0.020 and [Z]\,=\,0.014.}
         \label{metallicity_diff}
   \end{figure}

\subsection{Astrophysical parameters} \label{astrophysical_parameters}

As next step, we determined and analysed the astrophysical parameters of our sample stars, that is, the effective temperature and luminosity, from which we deduced the evolutionary status. 

As our sample stars are members of the Galactic disk and situated beyond 1000\,pc from the Sun, interstellar reddening (absorption) needs to be taken into account. 
Unfortunately, Str{\"o}mgren-Crawford indices, which allow a reliable reddening estimation, are not available for our sample stars \citep{2015A&A...580A..23P}. We therefore relied on the reddening map of \citet{green2019} as well as the Gaia EDR3 photogeometric distances and their errors from \citet{2021AJ....161..147B} in order to interpolate within this map.

To derive the $T_{\rm eff}$ values, we used the photometric data from 2MASS \citep{2006AJ....131.1163S}, APASS \citep{2014CoSka..43..518H}, and Gaia DR2 \citep{gaia_dr2}. Then, several independent calibrations were applied, as listed in
\citet{2006A&A...458..293P,2017BlgAJ..26...45P,2008A&A...491..545N}. Furthermore, all available photometric data (ultraviolet, optical and near-infrared bands) were employed to analyse the spectral energy distribution (SED) of our sample stars. For the SED fitting, the VO SED Analyzer was used \citep[VOSA]{2008A&A...492..277B}, 
which is a browser-based online tool. The photometric data were derived from the available VO catalogues. Two different stellar atmosphere models were used to determine the effective temperatures: the blackbody model and the Kurucz ATLAS9 model \citep{Kurucz} using new opacity distribution functions \citep{2003IAUS..210P.A20C}
and solar metallicity.
The derived values are generally in good agreement with the photometrically calibrated ones. The errors
were derived by calculating the standard deviations from all available individual $T_{\rm eff}$ values.

The luminosity of our target stars was calculated using the photogeometric distances and their errors (directly
transformed to the error for the luminosity) from \citet{2021AJ....161..147B} and the bolometric corrections from \citet{1996ApJ...469..355F}. 
ZTFJ232952.58+501457.3 was excluded from further analysis because of its large luminosity uncertainty.

Ages, masses, and evolutionary phases were estimated by the Stellar Isochrone Fitting Tool\footnote{\url{https://github.com/Johaney-s/StIFT}}, which follows the methods described by \citet{malkov10}. PARSEC isochrones \citep{2012MNRAS.427..127B} for a solar metallicity of [Z]\,=\,0.020 were used in this context. We favour this value because it has been shown to be consistent with most recent results of Helioseismology \citep{2019BSRSL..88...50B,2019Atoms...7...41V}.

In Fig. \ref{histograms}, the distributions of the $T_{\rm eff}$, mass and evolutionary phase values are
shown. The latter are taken from the evolutionary models by \citet{2012MNRAS.427..127B,2014MNRAS.444.2525C}.
They divided their tracks and isochrones in fifteen phases from the pre-main sequence to the thermally pulsing AGB
stage, respectively.
In line with the spectral classifications presented in Sect. \ref{spectral_classification}, most of the stars have $T_{\rm eff}$ values between 7300 and 8500\,K (spectral types between F0 and A3). The masses are also consistent with the derived spectral types. Except for a single star (ZTFJ071633.43-020221.4), all objects are located between the zero-age main sequence (ZAMS, phase 5) and the terminal-age main 
sequence (TAMS, phase 6). Therefore, they are all hydrogen-burning main-sequence objects, which is additional evidence that the catalogue classification as RS CVn stars is clearly wrong \citep{2020ApJS..249...18C}.

A recent article by \citet{2021arXiv210501661A} estimated the solar metallicity to [Z]\,=\,0.014, which is significantly lower than the value used in this work. We have therefore recalculated the corresponding values using [Z]\,=\,0.014. A comparison of the results based on the two different metallicities is provided in Fig.~\ref{metallicity_diff}. The mass differences between [Z]\,=\,0.020 and [Z]\,=\,0.014 amount on average to about 0.1\,M\textsubscript{\(\odot\)}, in the sense that a lower metallicity results 
in lower masses. In addition to that, adopting the lower metallicity results in a 10\,\% increase in the derived ages. From a statistical point of view, these differences are neglectable.
   
   \begin{figure}
   \centering
   \includegraphics[width=\hsize]{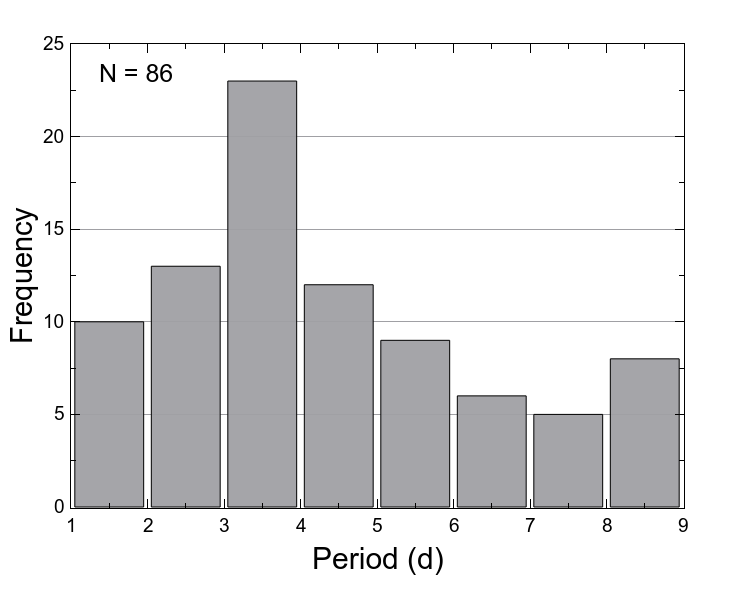}
      \caption{Distribution of periods among our target star sample, which reflects the selection process as described in Sect. \ref{data_selection}.}
         \label{periods}
   \end{figure}
   
\subsection{Stars in open cluster fields} \label{ocl_fields}

Due to the fact that the ZTF survey also targets the Galactic disk, we have searched for possible open cluster members among our sample stars, following the approach presented in \citet{2021A&A...645A..13P}. To this end, the angular distance from a given star to the cluster center together with proper motion and parallax constraints were applied. For the cluster parameters, the catalogues of \citet{2020A&A...640A...1C} and \citet{2021MNRAS.504..356D} were consulted. The astrometric data for our target star sample were taken from the Gaia EDR3 \citep{2021A&A...649A...1G}.

We only found six matches with the newly discovered open clusters of \citet[][``sc'' numbering scheme]{2019ApJS..245...32L} and SAI 43 \citep{2010AstL...36...75G}.
These are ZTFJ014156.09+604451.3 (SC 1332), ZTFJ050755.19+495727.2 (SAI 43), 
ZTFJ060420.41+265942.2 (SC 1374), ZTFJ062801.26+162502.0 (SC 1381),
ZTFJ064847.25+020447.8 (SC 1392 and SC 1393), and ZTFJ071633.43-020221.4 (SC 1977). In the case of ZTFJ064847.25+020447.8, we are unable to decide to which of the two clusters it belongs because the clusters are quite close together and have almost identical parameters. Comparing ages from our estimation (Section \ref{astrophysical_parameters}) with the corresponding cluster ages, we find excellent agreement. However, we emphasize that the age estimates for the clusters from \citet{2019ApJS..245...32L} exhibit quite large errors.

\begin{figure*}
   \centering
	 \begin{tabular}{cc}
	\subfigure{\includegraphics[width=0.47\textwidth]{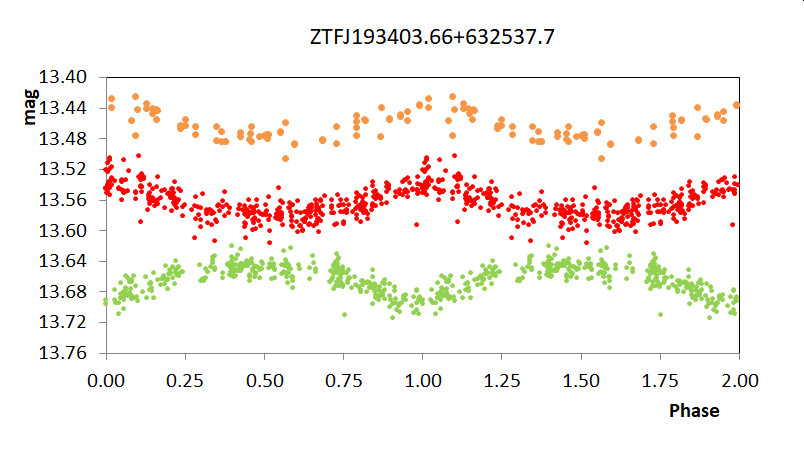}} &
    \subfigure{\includegraphics[width=0.47\textwidth]{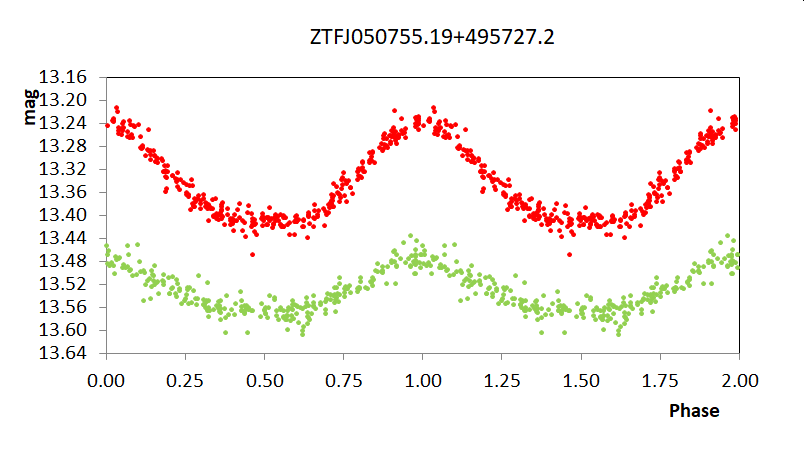}} \\
    \subfigure{\includegraphics[width=0.47\textwidth]{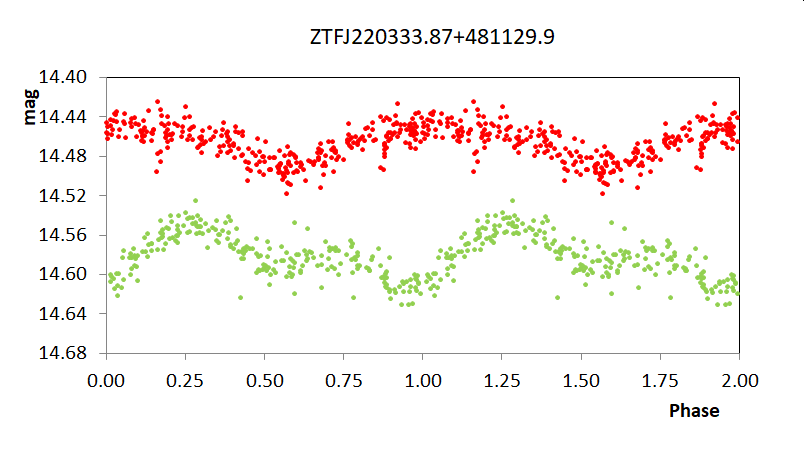}} &
    \subfigure{\includegraphics[width=0.47\textwidth]{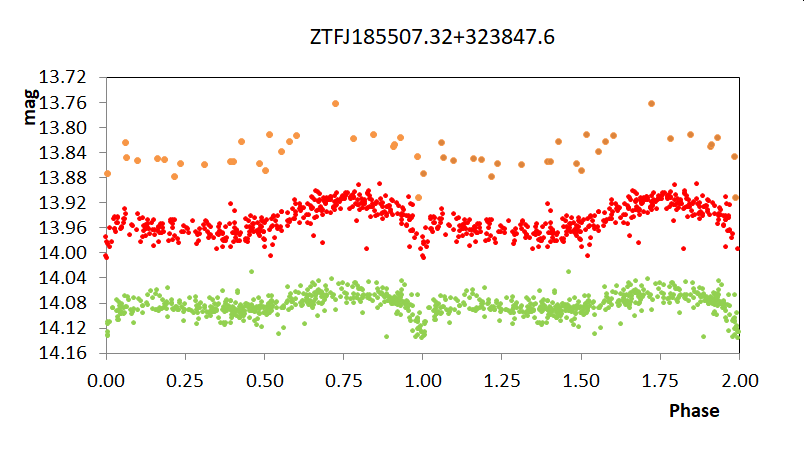}} \\
	 \end{tabular}
   \caption{Example light curves illustrating the four different subgroups among our sample stars: (i) $g$ and $r$ data are in anti-phase (upper left panel), (ii) the amplitude in $r$ is larger than in $g$ (upper right panel), (iii) the light curves in $g$ and $r$ are not consistent (lower left panel), and (iv) eclipses with possible additional ACV variability (lower right panel). Orange, red, and green symbols correspond to $i$, $r$, and $g$ band data, respectively.}
    \label{light_curves_selected}
\end{figure*}

\subsection{Light curves} \label{light_curves}

The histogram of the periods (Fig. \ref{periods}) reflects the selection process as described in
Sect. \ref{data_selection}. Although ACV variables can have periods of decades and even longer
\citep{2019IAUS..339..106M}, the detection of periods longer than ten days is hindered by gaps in the data and the rather low variability amplitude \citep{2016AJ....152..104H}.

A note is due on the variability of the non-magnetic CP1 and CP3 stars. The CP1 stars
are located in the classical pulsation strip and are known to exhibit $\gamma$ Doradus and $\delta$ Scuti variability
\citep{2017MNRAS.465.2662S}. Such variables were already excluded in the selection process 
(Section \ref{data_selection}). Both groups show also indications of rotational variability, but with amplitudes of
a few mmags only \citep{2015MNRAS.448.1378B,2021A&A...645A..34P,2021arXiv210709096K}. The amplitudes of
our variable star sample are much larger (Fig. \ref{light_curves_all}), which excludes a possible contamination with
CP1 and CP3 stars, respectively.

In general, concerning the characteristics of the light curves, our target star sample can be divided into four subgroups:

\begin{enumerate}
    \item $g$ and $r$ data are in anti-phase,
    \item the amplitude in $r$ is larger than in $g$ and the light curves are in phase,
    \item the light curves in $g$ and $r$ are not consistent (i.e. they show different shapes), 
    \item eclipses with possible additional ACV variability.
\end{enumerate}
In Fig. \ref{light_curves_selected} we present one typical example of each of these subgroups. The full set of light curves is shown in Fig. \ref{light_curves_all} in the appendix.

\citet{1973ApJ...179..527M,1975AJ.....80..137M} reported anti-phase variations between the far-ultraviolet and optical spectral regions suggesting that the global flux redistribution caused by the phase-dependent absorption is the probable cause of the observed phenomena. If so, there is a point or
a region in the spectrum where the flux remains almost unchanged (called the ``null wavelength region'') and varies more or less in anti-phase on both sides of it \citep{2007pms..conf..300M}. The first subgroup of our sample stars (anti-phase variations in $g$ and $r$), however, is unusual in that it shows anti-phase variations in the optical region, where, in general, the light changes are in phase in different photometric passbands \citep{2011A&A...525A..16P}. The null wavelength region of our ACV variables has to be somewhere between 4750 and 6200\,\AA, the central wavelengths of the $g$ and $r$ filters. One might speculate that the observed phenomenon is connected to the 5200\,\AA\ flux depression. ACV variables with anti-phase variations in the optical region are extremely rare \citep{2017NewA...50..104G}; therefore these stars are of special interest. Furthermore, it is obvious that this kind of anti-phase variations is a unique characteristic of ACV stars and very much suited to the identification of this particular group of stars in massive photometric time-series databases that contain observations in different passbands.

Due to a general lack of observations in the red optical region and the near-infrared \citep{1998A&AS..131...63C,2011mast.conf..476J}, the characteristics of the second subgroup (amplitude in $r$ is larger than in $g$) are difficult to investigate. The ZTF survey will help analyse this phenomena in more detail, which may turn out to be helpful in the detection of ACV variables because pulsating variables, for example, are expected to show larger amplitudes at shorter wavelengths. However, we have to emphasize that the light 
curves are in phase which is the most common case for ACV variables \citep{2011A&A...525A..16P}.

Different light curve characteristics in different filters and wavelength regions (third subgroup) has been well documented for ACV variables \citep{1986A&AS...64....9M}. Up to now, this phenomenon has almost exclusively been investigated in the narrow-band Str{\"o}mgren photometric system, which covers the spectrum up to 5700\,\AA. The ZTF data allows to investigate the light curve behaviour of ACV variables at even longer wavelengths, which, as the different light curve shapes in the $r$ band demonstrate, opens up an interesting new window on this phenomenon. Additional data in the $i$ band, in particular, will be important to further investigate the light curve characteristics of this subgroup and the correlation with spectral features.

The published overall binary fraction among CP2/4 stars is ranging from 20 to 51\,\% \citep{2020CoSka..50..570P}. However, eclipsing binary systems with such components are extremely rare \citep{2018MNRAS.478.1749K,2019MNRAS.487.4230S}. We therefore call particular attention to the fourth subgroup of stars (eclipses with possible additional ACV variability). We caution, however, that detailed high resolution spectrosocpy is needed to shed more light on the nature of these objects.

In summary, the quality of the ZTF data is comparable to that of other photometric ground-based surveys like ASAS-3, KELT, MASCARA, or SuperWASP \citep{bernhard2015b,2020MNRAS.493.3293B}, while extending to much fainter magnitudes. The ZTF is therefore well suited to search for new ACV variables and an analysis of their light curve characteristics. 

\section{Conclusions} \label{conclusions}

The ZTF survey is regularly monitoring a large part of the sky in three passbands ($g$, $r$, and $i$) to the limiting magnitude of 20.5 mag. Observations also include the Galactic disk, which is important to the study of upper-main sequence stars. Extending to much fainter magnitudes, the ZTF is an excellent supplement to the TESS satellite mission.

In this paper, we present a first case study about the potential of the ZTF survey to find and investigate new ACV variables, which are upper-main sequence stars that show variability due to chemical surface spots. Employing selection criteria on the basis of known astrophysical and light curve properties of this group of stars, we have identified 86 candidate ACV variables. 15 of these candidates could be confirmed as mCP stars with spectra from LAMOST DR7, which proves the viability of our approach.

ZTF data come in three different passbands, which will allow to apply the Bayesian photometric imaging technique \citep{2010A&A...509A..43L,2015A&A...574A..57P} to a significant number of ACV variables in the future. This method allows the mapping of surface characteristics and hence the study of the chemical spots. In so far, it is similar to a spectroscopic Doppler-imaging analysis, but it is not a
tomographic method. The investigation of extensive sets of ground-based spectropolarimetric data for Doppler- and magnetic Doppler-imaging and photometric data has yielded an excellent correlation of the surface maps from photometric and spectroscopic data in terms of spot location and size. The more data are available in different filters, the more accurate the description
of the stellar surface can be conducted \citep{2020A&A...642A..64W}. 

Our search for ACV stars in the list of published variables from the ZTF survey has shown that ZTF data are very much suited for the study of this type of variability. However, the available automatic search and classification algorithms employed with the ZTF data do not include the relatively rare class of ACV variables. To reduce the contamination of other variability classes with ACV variables and to expand the sample size of these stars, it is imperative to be able to correctly identify these stars in survey data of different characteristics. By expanding the current sample of new ACV variables and candidates, we are intending to set up and calibrate a search algorithm to identify these stars, which can then also be applied to any other set of photometric times series. Anti-phase variations, in particular, seem to be an interesting characteristic of a subset of ACV variables that is readily usable for the identification of ACV stars in massive photometric time-series databases with observations in different passbands.

\begin{acknowledgements}
The authors thank Sebasti\'{a}n Otero (VSX team) for helpful comments. EP acknowledges support by the Erasmus+ programme of the European Union under grant number 2020-1-CZ01-KA203-078200. Based on observations obtained with the Samuel Oschin 48-inch Telescope at the Palomar Observatory as part of the Zwicky Transient Facility project, which is supported by the National Science Foundation under Grant No. AST-1440341 and a collaboration including Caltech, IPAC, the Weizmann Institute for Science, the Oskar Klein Center at Stockholm University, the University of Maryland, the University of Washington, Deutsches Elektronen-Synchrotron and Humboldt University, Los Alamos National Laboratories, the TANGO Consortium of Taiwan, the University of Wisconsin at Milwaukee, and Lawrence Berkeley National Laboratories. Operations are conducted by COO, IPAC, and UW. Guoshoujing Telescope (the Large Sky Area Multi-Object Fiber Spectroscopic Telescope LAMOST) is a National Major Scientific Project built by the Chinese Academy of Sciences. Funding for the project has been provided by the National Development and Reform Commission. LAMOST is operated and managed by the National Astronomical Observatories, Chinese Academy of Sciences. This work has made use of data from the European Space Agency (ESA) mission {\it Gaia} (\url{https://www.cosmos.esa.int/gaia}), processed by the {\it Gaia} Data Processing and Analysis Consortium (DPAC, \url{https://www.cosmos.esa.int/web/gaia/dpac/consortium}). Funding for the DPAC has been provided by national institutions, in particular the institutions
participating in the {\it Gaia} Multilateral Agreement.
This publication makes use of data products from the Two Micron All Sky Survey, which is a joint project of the University of Massachusetts and the Infrared Processing and Analysis Center/California Institute of Technology, funded by the National Aeronautics and Space Administration and the National Science Foundation.

\end{acknowledgements}

\bibliographystyle{aa} 
\bibliography{paper_bibliography.bib}

\begin{appendix} 

\section{Essential data and light curves for our sample stars}

Table \ref{table_master1} lists essential data for our sample stars. It is organised as follows:

\begin{itemize}
\item Column 1: ZTF ID.
\item Column 2: Identifier from Gaia EDR3.
\item Column 3: Right ascension (J2000). Positional information was taken from GAIA EDR3. \citep{2021A&A...649A...1G}.
\item Column 4: Declination (J2000).
\item Column 5: Period from \citet{2020ApJS..249...18C}.
\item Column 6: $G$ magnitude (Gaia DR2).
\item Column 7: $G$ magnitude error (Gaia DR2).
\item Column 8: Parallax (Gaia EDR3).
\item Column 9: Parallax error.
\item Column 10: Mean effective temperature.
\item Column 11: Mean effective temperature error.
\item Column 12: Luminosity.
\item Column 13: Luminosity error.
\item Column 14: Mass.
\item Column 15: Evolutionary Phase.
\end{itemize}

\clearpage

\setcounter{table}{0}  
\begin{table*}
\caption{Essential data for our sample stars, sorted by increasing right ascension. The columns denote: 
(1) ZTF ID. (2) Gaia EDR3 ID (3) Right ascension (J2000; GAIA EDR3). (4) Declination (J2000; GAIA EDR3). 
(5) Period from \citet{2020ApJS..249...18C}. (6) $G$ magnitude (Gaia DR2). (7) $G$ magnitude error (Gaia DR2).
(8) Parallax (Gaia EDR3). (9) Parallax error.
(10) Mean effective temperature. (11) Mean effective temperature error.
(12) Luminosity. (13) Luminosity error.
(14) Mass. (15) Evolutionary Phase.}  
\label{table_master1}
\begin{center}
\begin{adjustbox}{max width=\textwidth,angle=90}
\begin{tabular}{lllcclcccccccccccccc}
\hline
\hline
(1) & (2) & (3) & (4) & (5) & (6) & (7) & (8) & (9) & (10) & (11) & (12) & (13) & (14) & (15) \\
ID\_ZTF	&	ID\_EDR3	&	RA(J2000) 	&	 Dec(J2000)    	&	Period	& $G$\,mag	&	e\_$G$\,mag	&
$\pi$ (EDR3)	&	e\_$\pi$ (EDR3) & $\log T_\mathrm{eff}$ &
e\_$\log T_\mathrm{eff}$ & $\log L/L_\odot$ & e\_$\log L/L_\odot$ & Mass & Phase \\
\hline
ZTFJ000520.70+604135.4	&	429317960188135296	&	1.3362618	&	60.6931614	&	5.211591	&	12.6493	&	0.0008	&	0.9322	&	0.0109	&	3.923	&	0.021	&	1.27	&	0.01	&	1.971	&	5.465	\\
ZTFJ001616.48+622426.9	&	431246572303647872	&	4.0687157	&	62.4074779	&	3.798443	&	13.5277	&	0.0017	&	0.4380	&	0.0180	&	4.099	&	0.034	&	2.16	&	0.05	&	3.354	&	5.350	\\
ZTFJ003646.27+602633.8	&	427090865026728960	&	9.1928387	&	60.4427282	&	1.978370	&	14.8742	&	0.0020	&	0.3004	&	0.0196	&	3.962	&	0.018	&	1.42	&	0.06	&	2.173	&	5.377	\\
ZTFJ005928.14+603315.8	&	426494959780476672	&	14.8672664	&	60.5543699	&	1.554077	&	14.6406	&	0.0019	&	0.2667	&	0.0203	&	3.966	&	0.017	&	1.67	&	0.07	&	2.400	&	5.651	\\
ZTFJ012145.50+443625.8	&	397603921663368704	&	20.4396315	&	44.6071471	&	5.155296	&	12.5811	&	0.0004	&	0.8321	&	0.0170	&	3.873	&	0.006	&	1.08	&	0.02	&	1.748	&	5.566	\\
ZTFJ013330.13+530809.5	&	407448399025891968	&	23.3756020	&	53.1359575	&	1.414782	&	15.0926	&	0.0008	&	0.2753	&	0.0276	&	3.907	&	0.015	&	1.24	&	0.09	&	1.923	&	5.535	\\
ZTFJ013347.59+532935.7	&	407478532516336512	&	23.4483379	&	53.4932444	&	3.205112	&	15.3200	&	0.0007	&	0.3245	&	0.0328	&	3.898	&	0.023	&	0.96	&	0.09	&	1.726	&	5.160	\\
ZTFJ014156.09+604451.3	&	509734591379435648	&	25.4837795	&	60.7475760	&	1.964292	&	14.6626	&	0.0015	&	0.3433	&	0.0193	&	3.996	&	0.040	&	1.62	&	0.13	&	2.430	&	5.394	\\
ZTFJ014923.64+593834.9	&	506595829275620864	&	27.3485254	&	59.6430339	&	8.084983	&	13.0248	&	0.0009	&	0.8180	&	0.0173	&	3.926	&	0.021	&	1.26	&	0.02	&	1.970	&	5.428	\\
ZTFJ015503.00+542632.7	&	408366491236405888	&	28.7625427	&	54.4424073	&	3.295617	&	15.1563	&	0.0005	&	0.2460	&	0.0328	&	3.872	&	0.015	&	1.17	&	0.12	&	1.817	&	5.658	\\
ZTFJ040517.81+335007.7	&	170866138481187712	&	61.3242246	&	33.8354789	&	3.227968	&	15.6196	&	0.0005	&	0.2778	&	0.0390	&	3.899	&	0.023	&	0.98	&	0.12	&	1.736	&	5.201	\\
ZTFJ050455.54+394433.7	&	200479761891366912	&	76.2314130	&	39.7426631	&	4.382893	&	14.5562	&	0.0011	&	0.4059	&	0.0233	&	3.904	&	0.008	&	1.26	&	0.08	&	1.933	&	5.577	\\
ZTFJ050714.45+405114.0	&	201061197382735616	&	76.8102274	&	40.8538710	&	2.064551	&	14.8037	&	0.0006	&	0.3951	&	0.0302	&	3.897	&	0.013	&	1.13	&	0.07	&	1.825	&	5.469	\\
ZTFJ050755.19+495727.2	&	262338771402441728	&	76.9799733	&	49.9575289	&	3.624936	&	13.3362	&	0.0033	&	0.3453	&	0.0151	&	3.989	&	0.007	&	1.99	&	0.04	&	2.803	&	5.795	\\
ZTFJ051331.04+393032.4	&	188685373675932288	&	78.3793882	&	39.5089795	&	3.312342	&	14.3473	&	0.0005	&	0.4542	&	0.0221	&	3.866	&	0.005	&	1.19	&	0.04	&	1.828	&	5.702	\\
ZTFJ052856.69+475711.7	&	212446919462212480	&	82.2362356	&	47.9532435	&	3.835257	&	14.0640	&	0.0008	&	0.4955	&	0.0184	&	3.887	&	0.010	&	1.14	&	0.03	&	1.814	&	5.547	\\
ZTFJ053725.01+382012.3	&	189705617387154560	&	84.3542173	&	38.3367281	&	3.462881	&	14.1013	&	0.0036	&	0.3561	&	0.0198	&	4.041	&	0.022	&	2.08	&	0.05	&	3.034	&	5.652	\\
ZTFJ054009.21+120945.1	&	3340374156920950144	&	85.0383933	&	12.1625140	&	3.056004	&	14.0169	&	0.0012	&	0.5193	&	0.0185	&	3.894	&	0.011	&	1.15	&	0.03	&	1.834	&	5.513	\\
ZTFJ054550.47+533903.3	&	264305625903932288	&	86.4603210	&	53.6508921	&	3.491721	&	14.2157	&	0.0030	&	0.2448	&	0.0188	&	3.946	&	0.032	&	1.64	&	0.07	&	2.337	&	5.712	\\
ZTFJ055527.76+152315.3	&	3347276684763140480	&	88.8656791	&	15.3875846	&	2.190848	&	12.9088	&	0.0009	&	0.7895	&	0.0214	&	3.904	&	0.016	&	1.13	&	0.02	&	1.837	&	5.417	\\
ZTFJ060420.41+265942.2	&	3430370966070322304	&	91.0850730	&	26.9950461	&	1.472744	&	13.4932	&	0.0043	&	0.4286	&	0.0153	&	4.058	&	0.009	&	1.85	&	0.03	&	2.852	&	5.206	\\
ZTFJ062241.93-011942.6	&	3118315482907229952	&	95.6747684	&	-1.3285408	&	2.827172	&	12.8498	&	0.0027	&	0.7891	&	0.0181	&	3.886	&	0.007	&	1.19	&	0.02	&	1.848	&	5.606	\\
ZTFJ062701.11+192226.7	&	3372391213970028544	&	96.7546478	&	19.3740622	&	4.398029	&	13.3396	&	0.0005	&	0.6161	&	0.0174	&	3.927	&	0.022	&	1.29	&	0.02	&	1.993	&	5.463	\\
ZTFJ062714.26+172512.9	&	3369750908593873408	&	96.8094451	&	17.4202275	&	1.138993	&	13.0612	&	0.0009	&	0.6967	&	0.0250	&	3.888	&	0.010	&	1.20	&	0.03	&	1.860	&	5.605	\\
ZTFJ062801.26+162502.0	&	3369362368672613632	&	97.0053128	&	16.4171893	&	3.614991	&	14.4531	&	0.0010	&	0.3072	&	0.0235	&	3.938	&	0.013	&	1.48	&	0.07	&	2.171	&	5.611	\\
ZTFJ063453.43+211828.5	&	3373098646622809216	&	98.7226575	&	21.3078995	&	6.061917	&	14.2772	&	0.0005	&	0.3716	&	0.0232	&	3.884	&	0.018	&	1.21	&	0.05	&	1.863	&	5.635	\\
ZTFJ063557.01+273821.5	&	3385992378961359104	&	98.9875744	&	27.6392864	&	3.326537	&	15.0112	&	0.0020	&	0.1924	&	0.0297	&	3.875	&	0.003	&	1.48	&	0.13	&	2.099	&	5.851	\\
ZTFJ064847.25+020447.8	&	3126354802693282176	&	102.1969031	&	2.0799216	&	1.039395	&	14.4483	&	0.0019	&	0.2742	&	0.0257	&	4.012	&	0.014	&	1.69	&	0.09	&	2.540	&	5.361	\\
ZTFJ065551.61+174248.2	&	3364299358142868736	&	103.9650861	&	17.7133839	&	7.763174	&	14.3983	&	0.0012	&	0.3558	&	0.0238	&	3.874	&	0.024	&	1.06	&	0.06	&	1.738	&	5.537	\\
ZTFJ065621.40+074658.6	&	3157081922142073216	&	104.0891923	&	7.7829449	&	8.939963	&	15.5032	&	0.0008	&	0.2376	&	0.0332	&	3.876	&	0.014	&	1.04	&	0.12	&	1.729	&	5.500	\\
ZTFJ065821.76-042555.4	&	3101702927363649664	&	104.5906796	&	-4.4320896	&	4.953162	&	13.0437	&	0.0027	&	0.5981	&	0.0265	&	3.889	&	0.015	&	1.32	&	0.05	&	1.958	&	5.704	\\
ZTFJ070329.94+030932.7	&	3115852576861974784	&	105.8747822	&	3.1590828	&	3.163433	&	13.0839	&	0.0010	&	0.7742	&	0.0142	&	3.869	&	0.012	&	0.97	&	0.02	&	1.671	&	5.459	\\
ZTFJ070334.53+080113.1	&	3154136815825083392	&	105.8939111	&	8.0203100	&	6.531035	&	14.1512	&	0.0019	&	0.3365	&	0.0213	&	3.881	&	0.021	&	1.22	&	0.05	&	1.868	&	5.658	\\
ZTFJ070618.89+010359.7	&	3114559551188271744	&	106.5787204	&	1.0665777	&	4.219654	&	15.6355	&	0.0012	&	0.2576	&	0.0381	&	3.865	&	0.012	&	0.97	&	0.13	&	1.664	&	5.487	\\
ZTFJ071633.43-020221.4	&	3108308758870686976	&	109.1393315	&	-2.0392960	&	3.286375	&	14.1493	&	0.0008	&	0.1306	&	0.0252	&	3.948	&	0.019	&	2.12	&	0.17	&	2.800	&	7.043	\\
ZTFJ072446.47-062829.6	&	3055096622329654016	&	111.1936605	&	-6.4749124	&	4.890556	&	14.8763	&	0.0010	&	0.3210	&	0.0244	&	3.880	&	0.010	&	1.00	&	0.07	&	1.706	&	5.417	\\
ZTFJ072604.40-062330.9	&	3055054394210945280	&	111.5183751	&	-6.3919508	&	8.429428	&	12.1658	&	0.0009	&	0.9867	&	0.0143	&	3.913	&	0.022	&	1.10	&	0.01	&	1.836	&	5.292	\\
ZTFJ182142.79+264259.6	&	4585174323482264960	&	275.4283113	&	26.7165481	&	2.826906	&	14.6153	&	0.0004	&	0.3318	&	0.0185	&	3.893	&	0.024	&	1.13	&	0.05	&	1.817	&	5.496	\\
ZTFJ182441.88+485404.5	&	2123491105988632448	&	276.1745200	&	48.9012394	&	4.529918	&	14.3043	&	0.0005	&	0.3527	&	0.0174	&	3.903	&	0.021	&	1.13	&	0.04	&	1.836	&	5.424	\\
ZTFJ183551.46+334749.7	&	2094333809969407488	&	278.9644468	&	33.7971106	&	2.521756	&	13.0377	&	0.0006	&	0.6470	&	0.0178	&	3.916	&	0.018	&	1.12	&	0.02	&	1.852	&	5.299	\\
ZTFJ185507.32+323847.6	&	2090413226384627200	&	283.7804988	&	32.6465520	&	1.927790	&	13.8901	&	0.0006	&	0.4885	&	0.0126	&	3.848	&	0.016	&	1.04	&	0.02	&	1.690	&	5.663	\\
ZTFJ185720.49+282654.0	&	2040346586458802176	&	284.3354104	&	28.4482882	&	4.297991	&	13.6940	&	0.0003	&	0.5476	&	0.0120	&	3.909	&	0.014	&	1.10	&	0.02	&	1.828	&	5.329	\\
ZTFJ190408.37+445439.3	&	2106330993656748288	&	286.0349001	&	44.9109022	&	3.544406	&	13.6395	&	0.0004	&	0.5222	&	0.0113	&	3.906	&	0.015	&	1.05	&	0.02	&	1.789	&	5.270	\\
ZTFJ190809.13+363121.0	&	2098738075595710848	&	287.0380757	&	36.5224814	&	7.244690	&	15.4943	&	0.0005	&	0.1511	&	0.0236	&	3.869	&	0.019	&	1.43	&	0.14	&	2.045	&	5.841	\\
ZTFJ190831.71+511959.8	&	2133769375046176768	&	287.1321686	&	51.3332604	&	4.488083	&	12.8144	&	0.0009	&	0.6689	&	0.0127	&	3.914	&	0.022	&	1.18	&	0.02	&	1.890	&	5.410	\\
ZTFJ192646.91+345654.2	&	2049748235512470912	&	291.6954743	&	34.9483612	&	3.353882	&	14.8057	&	0.0014	&	0.2615	&	0.0186	&	3.907	&	0.018	&	1.34	&	0.06	&	1.998	&	5.639	\\
ZTFJ192832.09+305205.8	&	2044736111779682176	&	292.1337101	&	30.8682693	&	6.858098	&	15.3430	&	0.0008	&	0.2686	&	0.0266	&	3.884	&	0.016	&	1.09	&	0.09	&	1.775	&	5.507	\\
ZTFJ193403.66+632537.7	&	2241889988402340736	&	293.5152836	&	63.4271347	&	3.040560	&	13.5126	&	0.0003	&	0.5245	&	0.0180	&	3.915	&	0.016	&	1.09	&	0.03	&	1.834	&	5.257	\\
ZTFJ193442.59+283840.2	&	2026247033497540992	&	293.6774920	&	28.6444965	&	1.126965	&	15.3178	&	0.0020	&	0.2534	&	0.0250	&	3.833	&	0.016	&	1.13	&	0.09	&	1.747	&	5.795	\\
ZTFJ193501.38+282225.7	&	2025483800652001152	&	293.7557711	&	28.3738018	&	3.187027	&	13.3942	&	0.0014	&	0.2221	&	0.0127	&	3.981	&	0.020	&	2.15	&	0.05	&	3.000	&	5.970	\\
ZTFJ193638.74+262513.5	&	2024971045912065536	&	294.1614368	&	26.4204089	&	2.040730	&	13.5042	&	0.0003	&	0.6374	&	0.0140	&	3.927	&	0.018	&	1.26	&	0.02	&	1.972	&	5.421	\\
ZTFJ194211.28+345344.4	&	2047363497889823616	&	295.5470250	&	34.8956649	&	8.357030	&	13.2659	&	0.0005	&	0.6911	&	0.0193	&	3.873	&	0.009	&	1.04	&	0.02	&	1.724	&	5.519	\\
ZTFJ194314.34+350354.7	&	2047360371153770240	&	295.8097744	&	35.0651743	&	4.579506	&	13.2115	&	0.0003	&	0.5074	&	0.0189	&	3.887	&	0.017	&	1.34	&	0.03	&	1.974	&	5.727	\\
ZTFJ194356.38+470830.9	&	2080554249216594944	&	295.9849448	&	47.1419201	&	5.374076	&	13.6845	&	0.0003	&	0.4765	&	0.0144	&	3.921	&	0.014	&	1.14	&	0.03	&	1.880	&	5.288	\\
ZTFJ194633.68+032219.9	&	4241823539654694272	&	296.6403529	&	3.3721901	&	7.804066	&	13.5785	&	0.0007	&	0.5930	&	0.0155	&	3.899	&	0.013	&	1.17	&	0.02	&	1.854	&	5.504	\\
ZTFJ194708.66+305907.6	&	2032227616496306048	&	296.7861319	&	30.9854390	&	2.639366	&	14.0226	&	0.0003	&	0.4938	&	0.0145	&	3.907	&	0.011	&	1.33	&	0.03	&	1.990	&	5.629	\\
ZTFJ195707.08+172944.9	&	1820520024151681792	&	299.2795278	&	17.4958011	&	7.053492	&	13.5815	&	0.0004	&	0.4439	&	0.0145	&	3.880	&	0.018	&	1.37	&	0.03	&	1.994	&	5.773	\\
ZTFJ195955.94+371754.2	&	2060214452368626176	&	299.9831171	&	37.2983694	&	8.501441	&	15.1303	&	0.0010	&	0.2645	&	0.0207	&	3.892	&	0.026	&	1.38	&	0.07	&	2.015	&	5.736	\\
ZTFJ200107.52+402140.8	&	2074295725974926208	&	300.2813808	&	40.3613362	&	3.020683	&	13.8560	&	0.0011	&	0.3612	&	0.0238	&	3.912	&	0.013	&	1.53	&	0.06	&	2.181	&	5.764	\\
ZTFJ200753.38+161817.2	&	1809400147648890240	&	301.9724288	&	16.3047641	&	6.998496	&	14.0547	&	0.0008	&	0.4379	&	0.0183	&	3.907	&	0.022	&	1.22	&	0.04	&	1.905	&	5.511	\\
ZTFJ201746.62+541552.6	&	2185170749074136448	&	304.4442574	&	54.2645952	&	2.892516	&	13.2046	&	0.0003	&	0.6501	&	0.0118	&	3.921	&	0.011	&	1.26	&	0.02	&	1.959	&	5.468	\\
ZTFJ203527.35+244945.9	&	1831531491338494080	&	308.8639635	&	24.8293879	&	5.174115	&	13.4936	&	0.0003	&	0.4827	&	0.0153	&	3.901	&	0.004	&	1.25	&	0.03	&	1.920	&	5.583	\\
ZTFJ203702.83+520647.9	&	2180824658793425280	&	309.2618259	&	52.1132953	&	8.610871	&	13.8723	&	0.0004	&	0.6258	&	0.0172	&	3.890	&	0.015	&	1.03	&	0.02	&	1.743	&	5.380	\\
ZTFJ203922.63+262122.2	&	1855754939389715328	&	309.8443026	&	26.3561266	&	3.852799	&	14.5034	&	0.0003	&	0.2758	&	0.0233	&	3.944	&	0.018	&	1.43	&	0.07	&	2.139	&	5.522	\\
ZTFJ205319.55+313746.4	&	1859630305566966912	&	313.3314857	&	31.6295237	&	3.559459	&	15.0831	&	0.0007	&	0.2840	&	0.0265	&	3.915	&	0.016	&	1.12	&	0.08	&	1.850	&	5.308	\\
ZTFJ211100.94+470048.1	&	2164472412637485056	&	317.7539826	&	47.0133465	&	2.523780	&	14.3300	&	0.0004	&	0.5850	&	0.0144	&	3.874	&	0.005	&	0.95	&	0.02	&	1.667	&	5.389	\\
ZTFJ212905.49+142755.5	&	1771272692152137088	&	322.2728965	&	14.4654079	&	4.655821	&	14.7020	&	0.0007	&	0.3010	&	0.0222	&	3.898	&	0.015	&	1.16	&	0.06	&	1.845	&	5.499	\\
ZTFJ213347.82+451647.7	&	1971144076219123584	&	323.4492766	&	45.2798944	&	8.467327	&	15.2336	&	0.0007	&	0.2678	&	0.0222	&	3.894	&	0.006	&	1.39	&	0.07	&	2.027	&	5.736	\\
ZTFJ213829.74+332758.7	&	1947187268337795328	&	324.6239446	&	33.4662933	&	5.704374	&	12.7055	&	0.0011	&	0.6050	&	0.0142	&	3.931	&	0.023	&	1.40	&	0.02	&	2.089	&	5.568	\\
ZTFJ214304.40+541759.0	&	2174297304939557888	&	325.7683695	&	54.2997118	&	2.204300	&	13.4431	&	0.0004	&	0.7522	&	0.0109	&	4.036	&	0.011	&	1.58	&	0.02	&	2.506	&	5.000	\\
ZTFJ215054.23+480704.0	&	1977830588687388160	&	327.7259888	&	48.1177538	&	6.024571	&	14.6211	&	0.0007	&	0.3914	&	0.0184	&	3.892	&	0.003	&	1.25	&	0.04	&	1.904	&	5.632	\\
ZTFJ215213.06+350116.1	&	1948994899811877504	&	328.0544517	&	35.0211370	&	6.551755	&	13.6075	&	0.0013	&	0.5333	&	0.0138	&	3.895	&	0.023	&	1.20	&	0.02	&	1.872	&	5.565	\\
ZTFJ215440.11+451202.7	&	1973564895885865728	&	328.6671634	&	45.2007511	&	2.237877	&	15.5960	&	0.0007	&	0.3017	&	0.0405	&	3.892	&	0.008	&	0.96	&	0.12	&	1.707	&	5.235	\\
ZTFJ220333.87+481129.9	&	1976145862000260736	&	330.8911355	&	48.1916256	&	3.319605	&	14.4741	&	0.0006	&	0.4437	&	0.0170	&	3.888	&	0.018	&	0.96	&	0.03	&	1.697	&	5.277	\\
ZTFJ220412.94+485912.5	&	1976602880880596480	&	331.0539267	&	48.9868040	&	3.797498	&	15.0597	&	0.0006	&	0.2436	&	0.0206	&	3.884	&	0.014	&	1.37	&	0.07	&	1.998	&	5.760	\\
ZTFJ221032.85+542437.5	&	2005454684479351680	&	332.6369071	&	54.4104017	&	4.285011	&	15.9310	&	0.0007	&	0.2432	&	0.0338	&	3.866	&	0.012	&	1.01	&	0.12	&	1.691	&	5.529	\\
\hline
\end{tabular}  
\end{adjustbox}  
\end{center}   
\end{table*}

\setcounter{table}{0}  
\begin{table*}
\caption{Essential data for our sample stars, sorted by increasing right ascension. The columns denote: 
(1) ZTF ID. (2) Gaia EDR3 ID (3) Right ascension (J2000; GAIA EDR3). (4) Declination (J2000; GAIA EDR3). 
(5) Period from \citet{2020ApJS..249...18C}. (6) $G$ magnitude (Gaia DR2). (7) $G$ magnitude error (Gaia DR2).
(8) Parallax (Gaia EDR3). (9) Parallax error.
(10) Mean effective temperature. (11) Mean effective temperature error.
(12) Luminosity. (13) Luminosity error.
(14) Mass. (15) Evolutionary Phase.}  
\label{table_master2}
\begin{center}
\begin{adjustbox}{max width=\textwidth,angle=90}
\begin{tabular}{lllcclcccccccccccccc}
\hline
\hline
(1) & (2) & (3) & (4) & (5) & (6) & (7) & (8) & (9) & (10) & (11) & (12) & (13) & (14) & (15) \\
ID\_ZTF	&	ID\_EDR3	&	RA(J2000) 	&	 Dec(J2000)    	&	Period	& $G$\,mag	&	e\_$G$\,mag	&
$\pi$ (EDR3)	&	e\_$\pi$ (EDR3) & $\log T_\mathrm{eff}$ &
e\_$\log T_\mathrm{eff}$ & $\log L/L_\odot$ & e\_$\log L/L_\odot$ & Mass & Phase \\
\hline
ZTFJ221304.21+533624.2	&	2004607162854210304	&	333.2675925	&	53.6067245	&	5.554406	&	15.4930	&	0.0021	&	0.1400	&	0.0302	&	3.949	&	0.019	&	1.95	&	0.19	&	2.704	&	5.882	\\
ZTFJ222357.65+474904.5	&	1987334835917530752	&	335.9902188	&	47.8178843	&	5.097425	&	16.0132	&	0.0007	&	0.1216	&	0.0327	&	3.902	&	0.010	&	1.45	&	0.23	&	2.092	&	5.748	\\
ZTFJ225444.50+572009.4	&	2010105446854516096	&	343.6854340	&	57.3359641	&	2.682164	&	13.0778	&	0.0013	&	0.5345	&	0.0543	&	3.958	&	0.007	&	1.57	&	0.09	&	2.290	&	5.592	\\
ZTFJ225454.64+542804.4	&	2002652269845315328	&	343.7276850	&	54.4678760	&	2.199670	&	13.2038	&	0.0032	&	0.4061	&	0.0129	&	4.055	&	0.023	&	2.06	&	0.03	&	3.057	&	5.547	\\
ZTFJ230203.90+535218.1	&	1996509951213474688	&	345.5162916	&	53.8716968	&	5.797225	&	14.7657	&	0.0014	&	0.3245	&	0.0215	&	3.919	&	0.024	&	1.29	&	0.06	&	1.978	&	5.517	\\
ZTFJ232236.78+512549.1	&	1991543109289628672	&	350.6532833	&	51.4303083	&	4.658976	&	12.5543	&	0.0029	&	0.4921	&	0.0131	&	4.016	&	0.017	&	1.87	&	0.02	&	2.728	&	5.569	\\
ZTFJ232952.58+501457.3	&	1943202500757860736	&	352.4691048	&	50.2492446	&	8.603556	&	16.5588	&	0.0016	&	0.0136	&	0.0543	&	3.880	&	0.037	&	3.14	&	3.47	&	5.087	&	7.712	\\
ZTFJ234424.36+583531.0	&	1999341159296500096	&	356.1015203	&	58.5919399	&	1.813400	&	14.0961	&	0.0007	&	0.3517	&	0.0176	&	3.941	&	0.012	&	1.71	&	0.04	&	2.402	&	5.782	\\
ZTFJ234442.88+512735.0	&	1944637191635780736	&	356.1786717	&	51.4597100	&	5.498045	&	16.2571	&	0.0012	&	0.1365	&	0.0378	&	3.913	&	0.043	&	1.33	&	0.24	&	1.999	&	5.599	\\
ZTFJ235154.72+621521.9	&	2012988125824883200	&	357.9780206	&	62.2560817	&	7.425652	&	14.1324	&	0.0006	&	0.3240	&	0.0150	&	4.057	&	0.048	&	2.11	&	0.04	&	3.123	&	5.589	\\
\hline
\end{tabular}  
\end{adjustbox}                                                                                                     
\end{center}                                                                                                        
\end{table*}

\begin{figure*}
   \centering
	 \begin{tabular}{ccc}
	\subfigure{\includegraphics[width=0.31\textwidth]{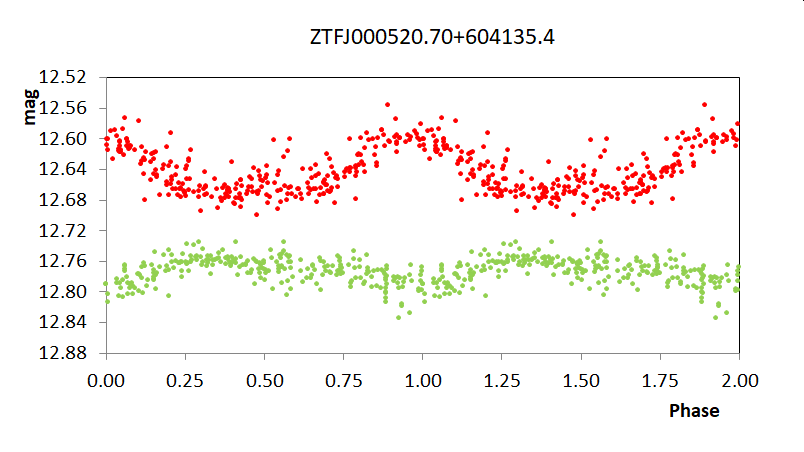}} &
    \subfigure{\includegraphics[width=0.31\textwidth]{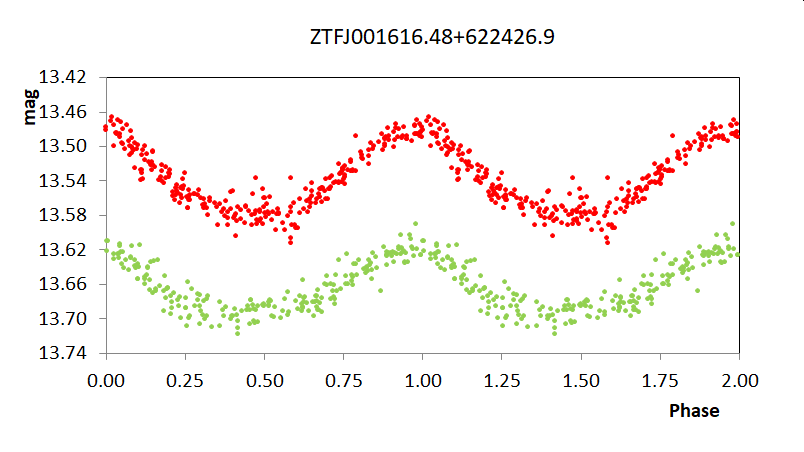}} &
    \subfigure{\includegraphics[width=0.31\textwidth]{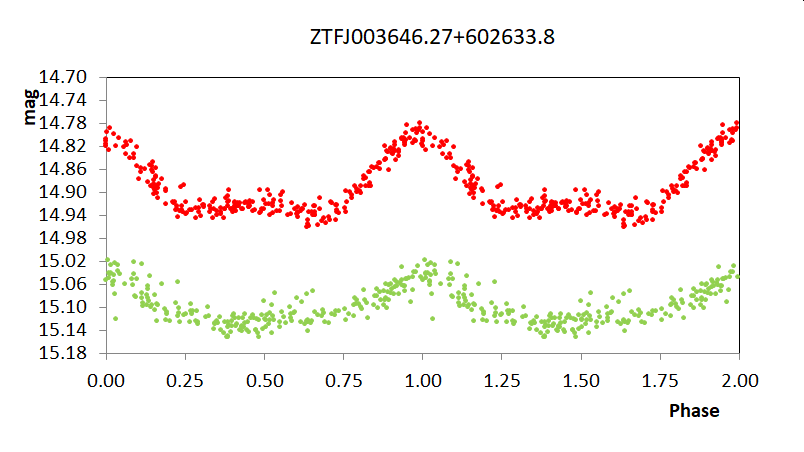}} \\
    \subfigure{\includegraphics[width=0.31\textwidth]{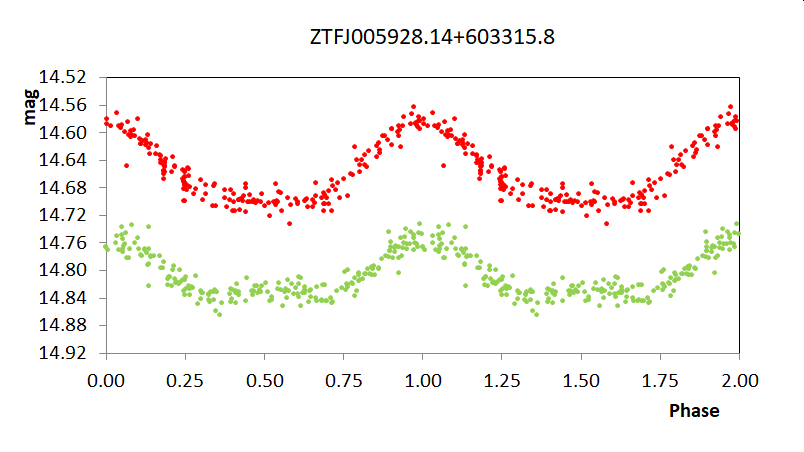}} &
    \subfigure{\includegraphics[width=0.31\textwidth]{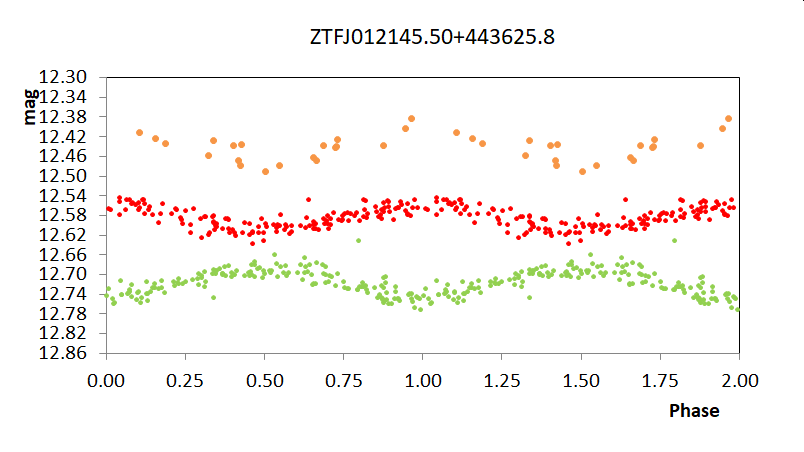}} &
    \subfigure{\includegraphics[width=0.31\textwidth]{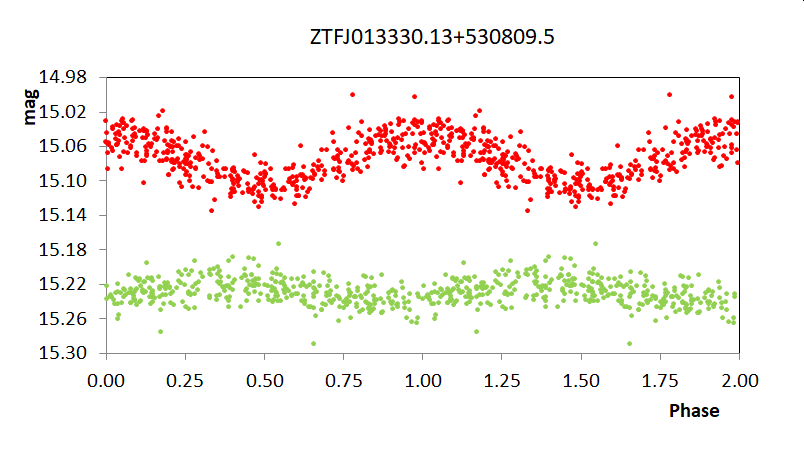}} \\
    \subfigure{\includegraphics[width=0.31\textwidth]{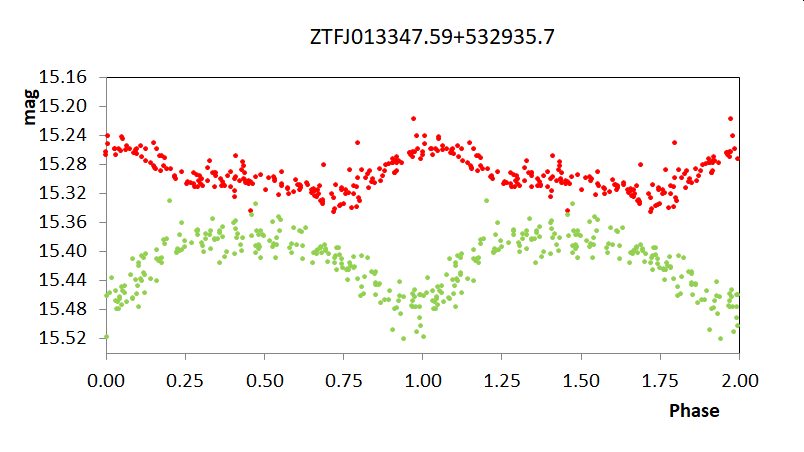}} &
    \subfigure{\includegraphics[width=0.31\textwidth]{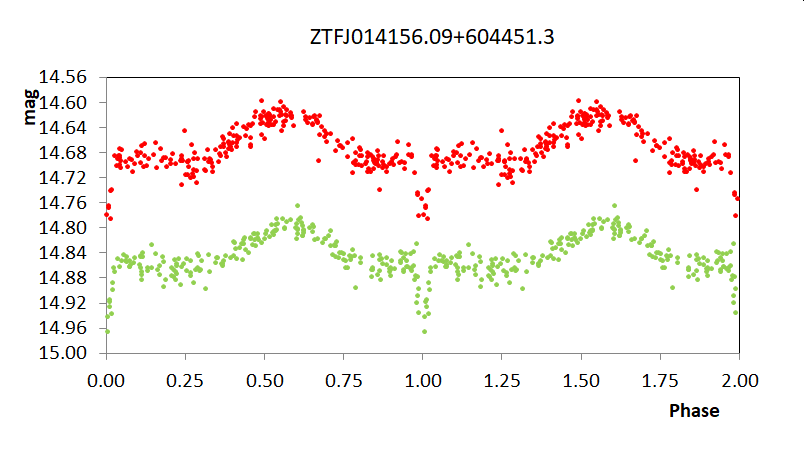}} &
    \subfigure{\includegraphics[width=0.31\textwidth]{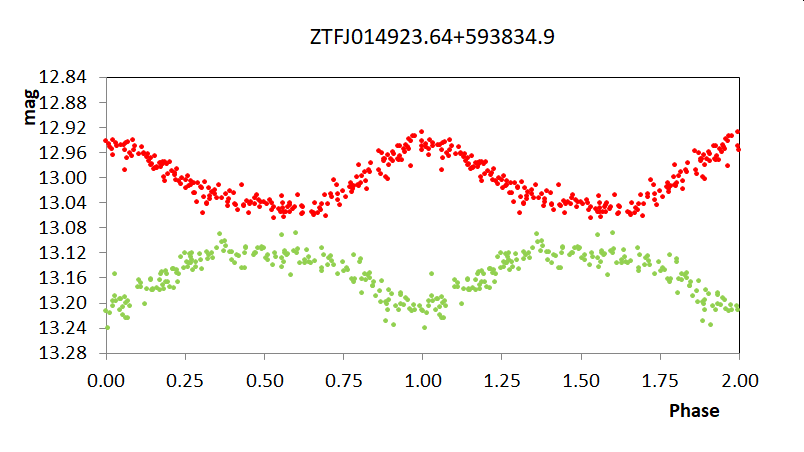}} \\
    \subfigure{\includegraphics[width=0.31\textwidth]{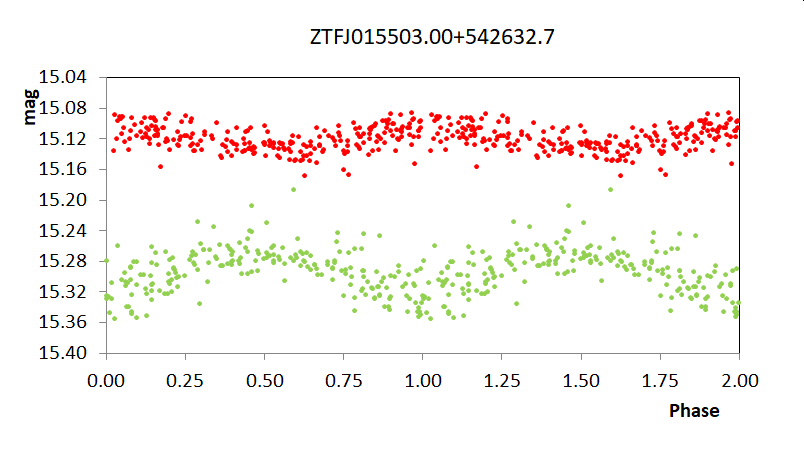}} &
    \subfigure{\includegraphics[width=0.31\textwidth]{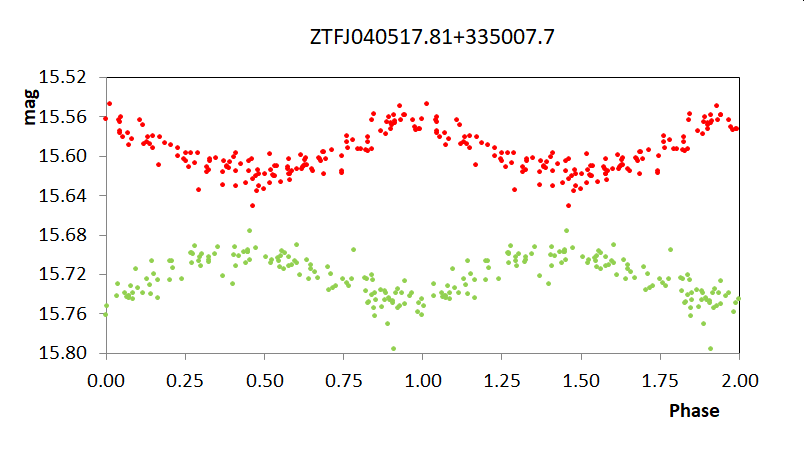}} &
    \subfigure{\includegraphics[width=0.31\textwidth]{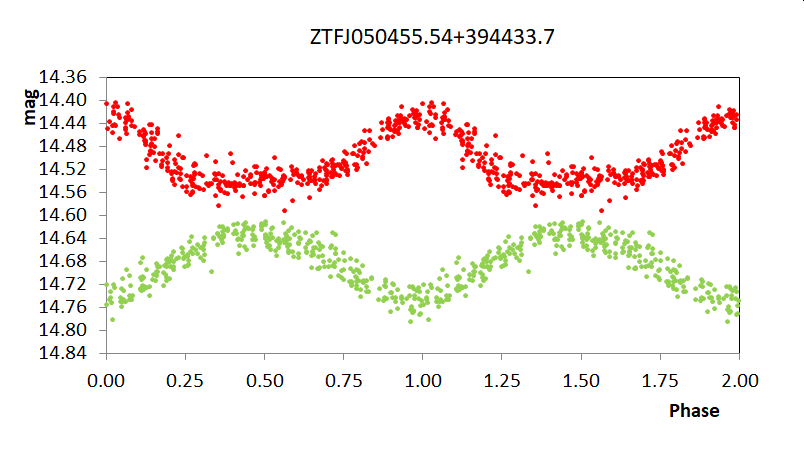}} \\
    \subfigure{\includegraphics[width=0.31\textwidth]{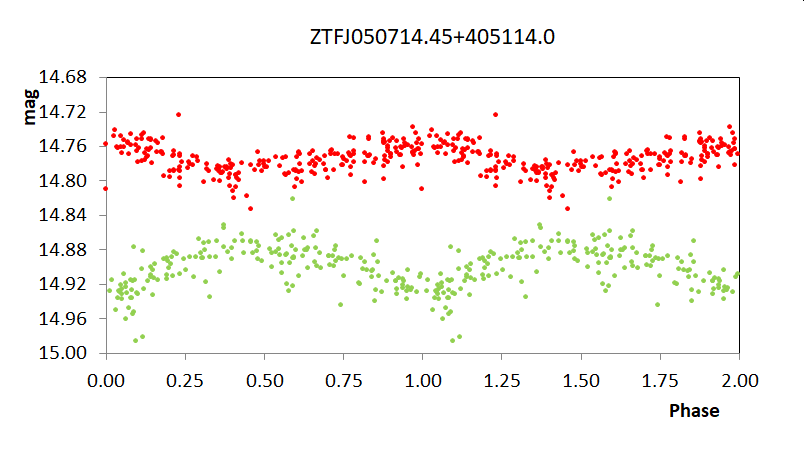}} &
    \subfigure{\includegraphics[width=0.31\textwidth]{14.png}} &
    \subfigure{\includegraphics[width=0.31\textwidth]{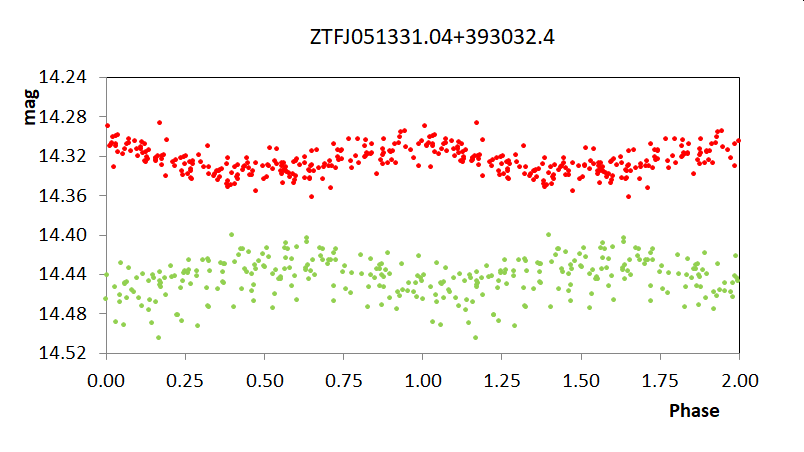}} \\
    \subfigure{\includegraphics[width=0.31\textwidth]{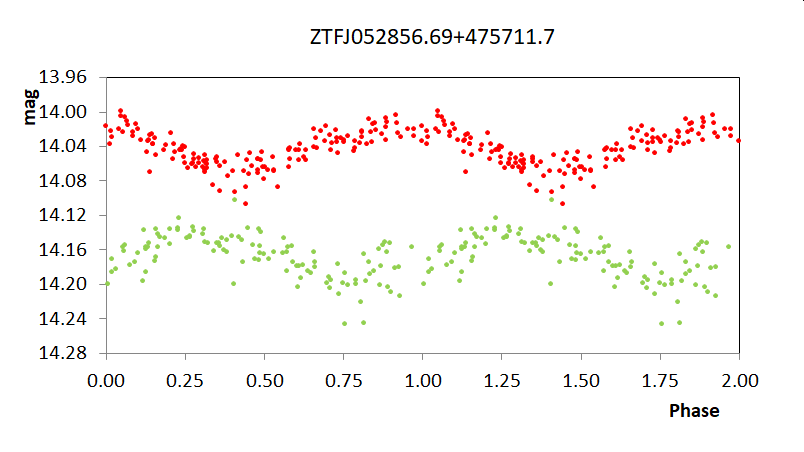}} &
    \subfigure{\includegraphics[width=0.31\textwidth]{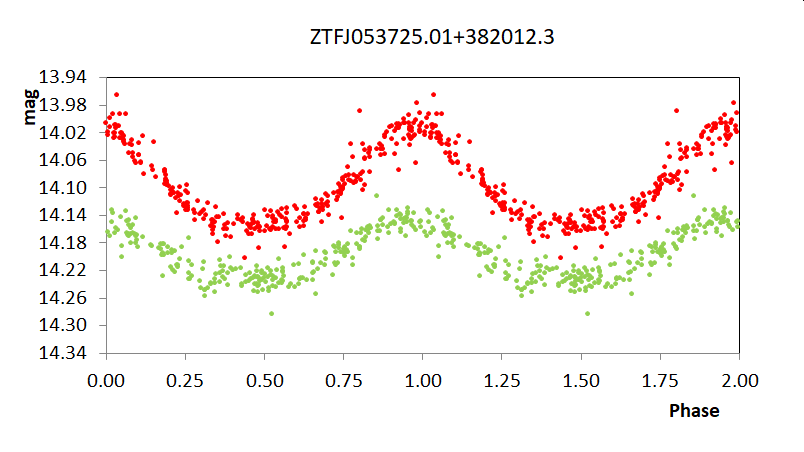}} &
    \subfigure{\includegraphics[width=0.31\textwidth]{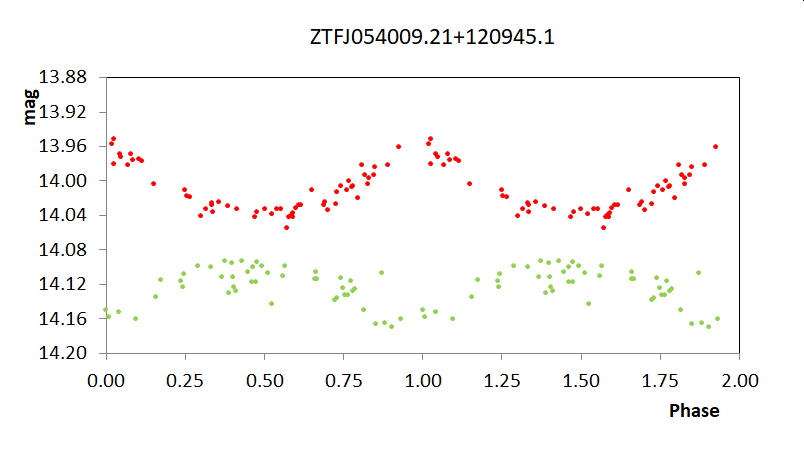}} \\
    \subfigure{\includegraphics[width=0.31\textwidth]{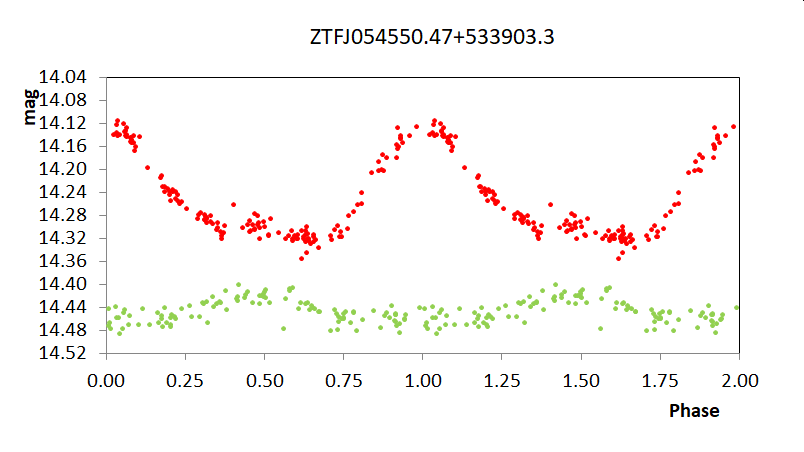}} &
    \subfigure{\includegraphics[width=0.31\textwidth]{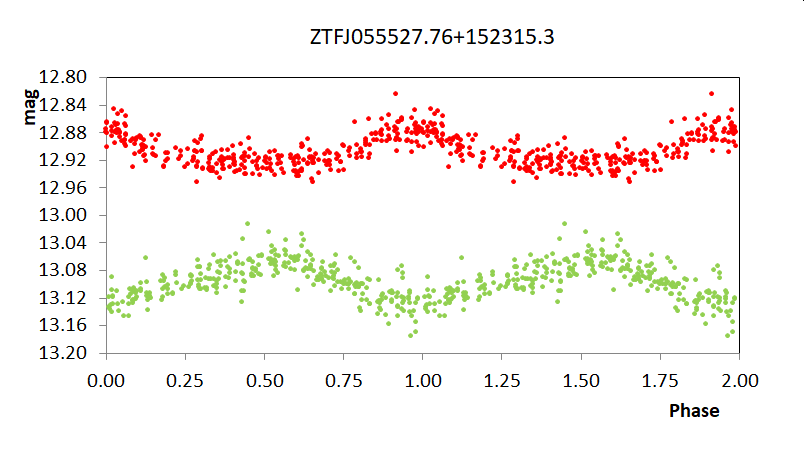}} &
    \subfigure{\includegraphics[width=0.31\textwidth]{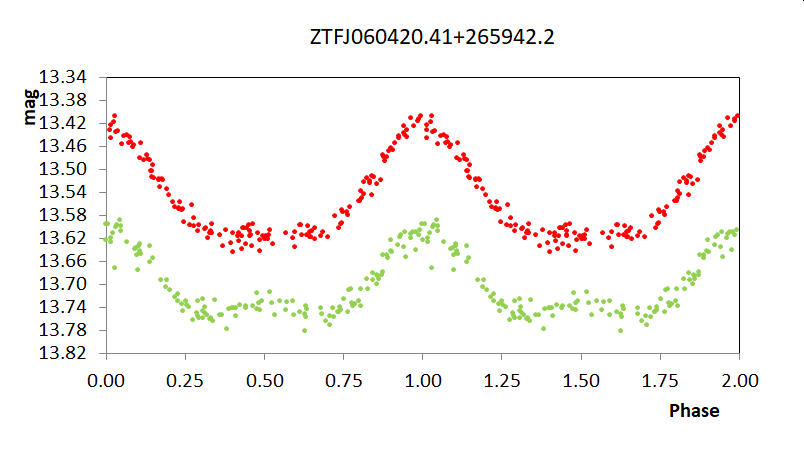}} \\
	 \end{tabular}
   \caption{Phased light curves of our target stars in all available passbands. Orange, red, and green symbols correspond to $i$, $r$, and $g$ band data, respectively.}
    \label{light_curves_all}
\end{figure*}
\addtocounter{figure}{-1}
\begin{figure*}
   \centering
	 \begin{tabular}{ccc}
	\subfigure{\includegraphics[width=0.31\textwidth]{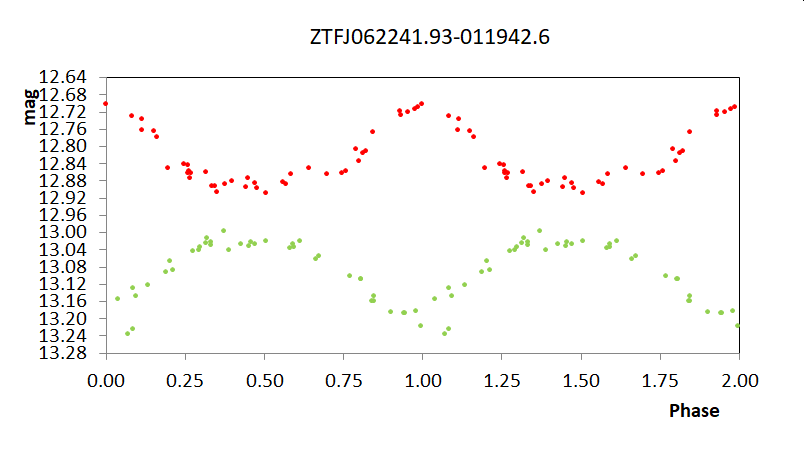}} &
    \subfigure{\includegraphics[width=0.31\textwidth]{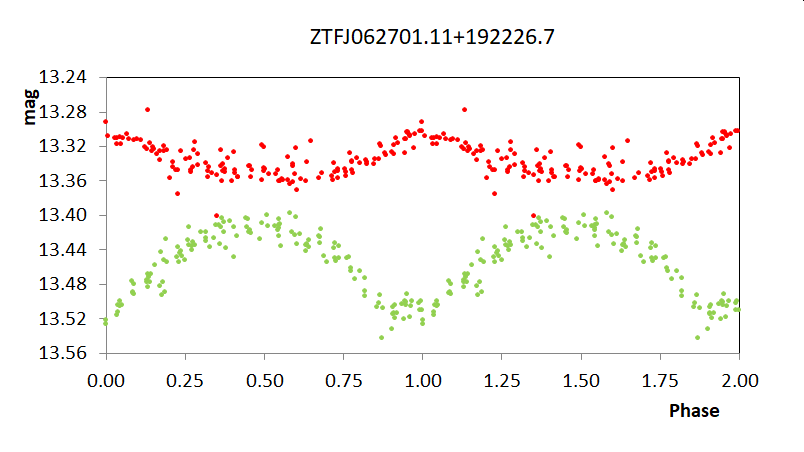}} &
    \subfigure{\includegraphics[width=0.31\textwidth]{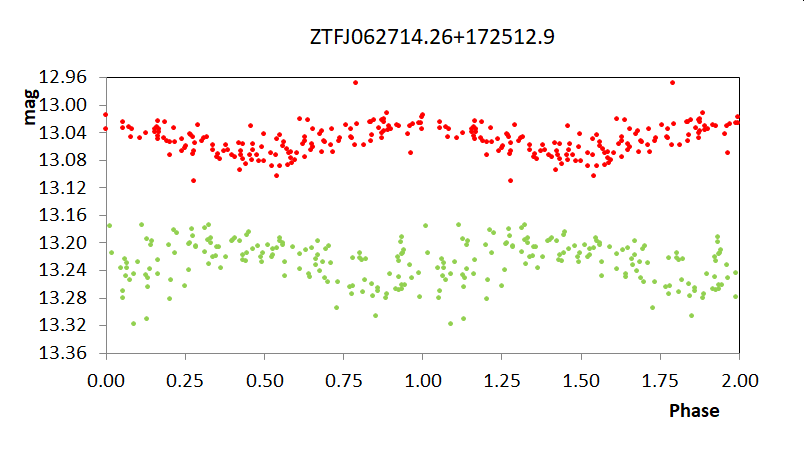}} \\
    \subfigure{\includegraphics[width=0.31\textwidth]{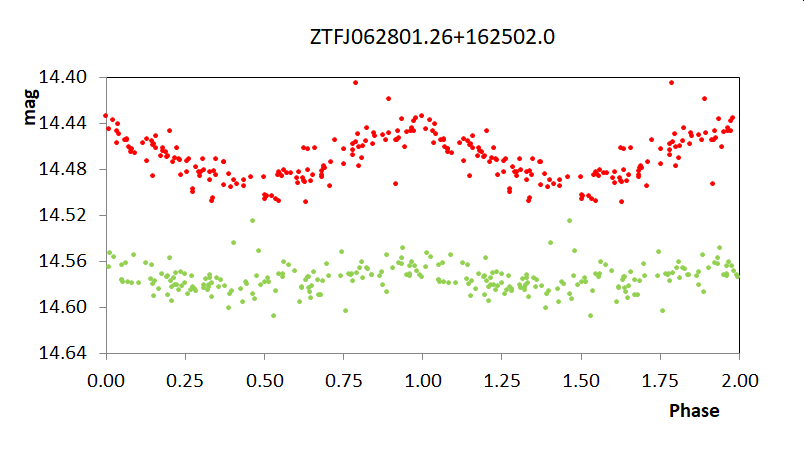}} &
    \subfigure{\includegraphics[width=0.31\textwidth]{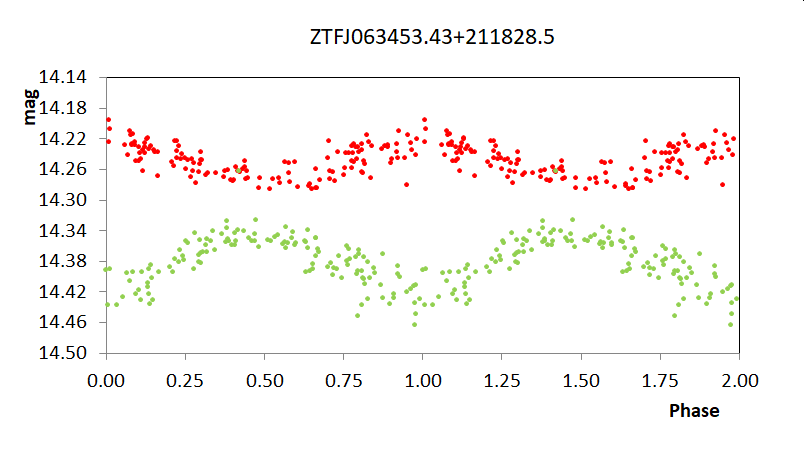}} &
    \subfigure{\includegraphics[width=0.31\textwidth]{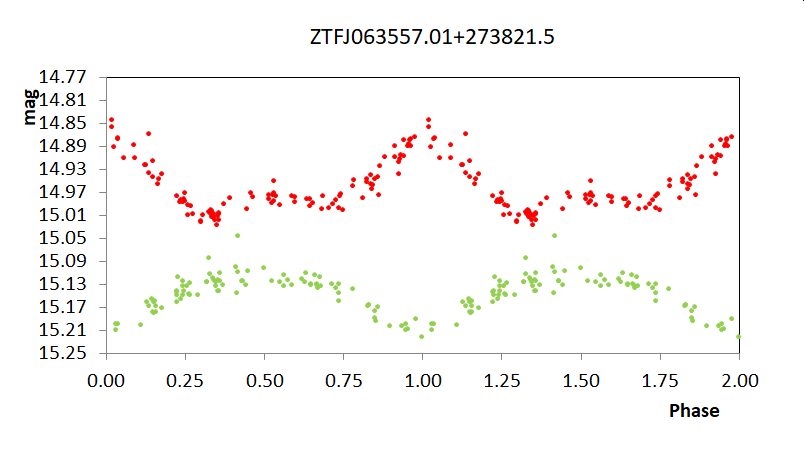}} \\
    \subfigure{\includegraphics[width=0.31\textwidth]{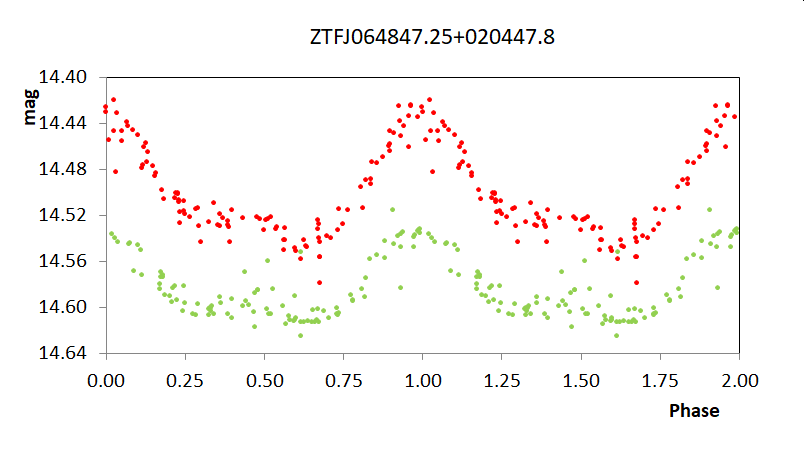}} &
    \subfigure{\includegraphics[width=0.31\textwidth]{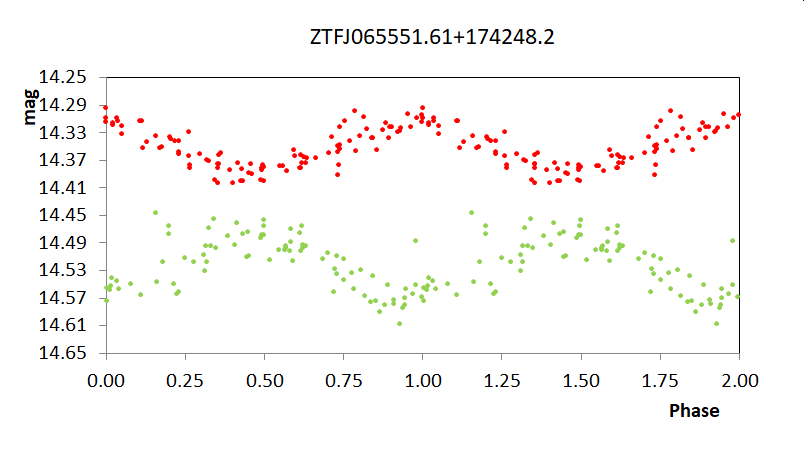}} &
    \subfigure{\includegraphics[width=0.31\textwidth]{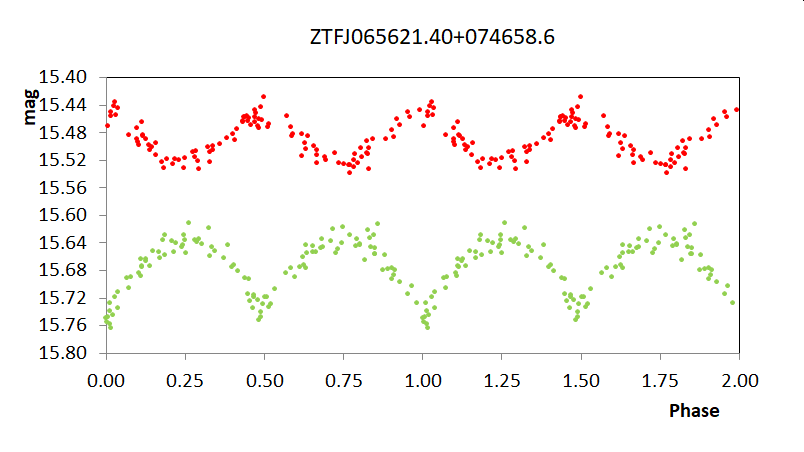}} \\
    \subfigure{\includegraphics[width=0.31\textwidth]{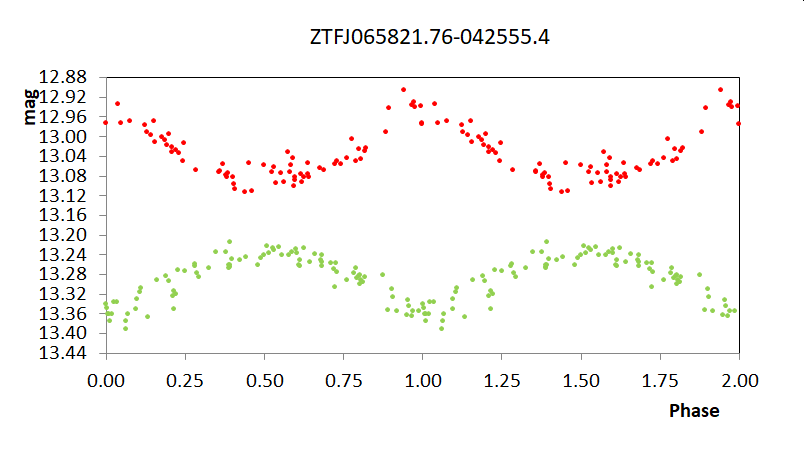}} &
    \subfigure{\includegraphics[width=0.31\textwidth]{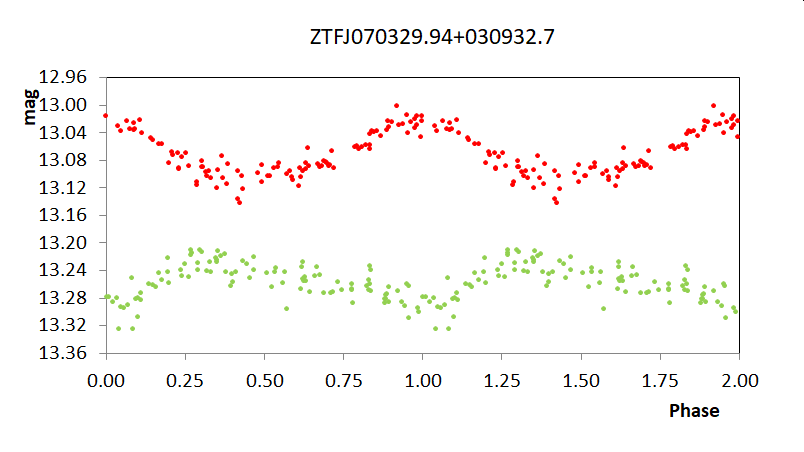}} &
    \subfigure{\includegraphics[width=0.31\textwidth]{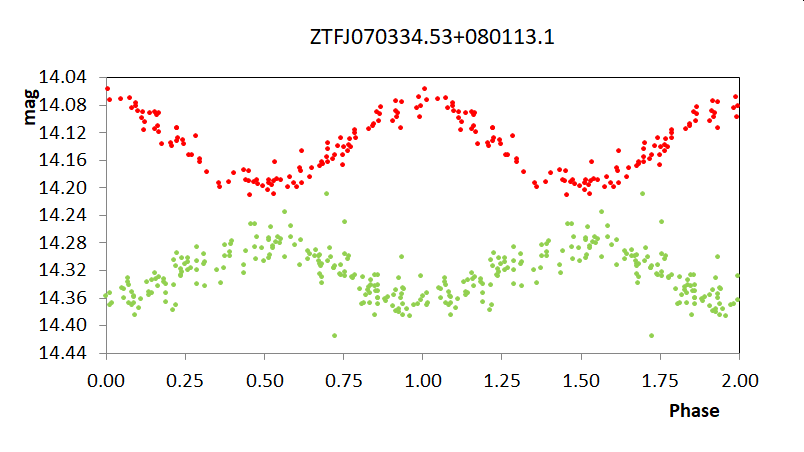}} \\
    \subfigure{\includegraphics[width=0.31\textwidth]{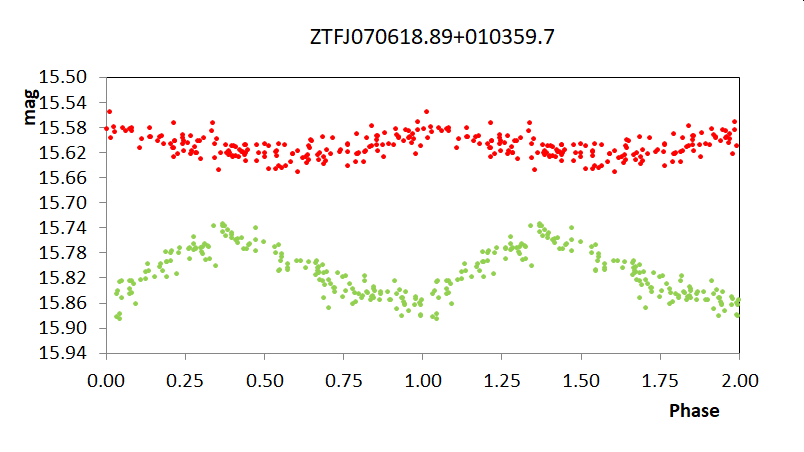}} &
    \subfigure{\includegraphics[width=0.31\textwidth]{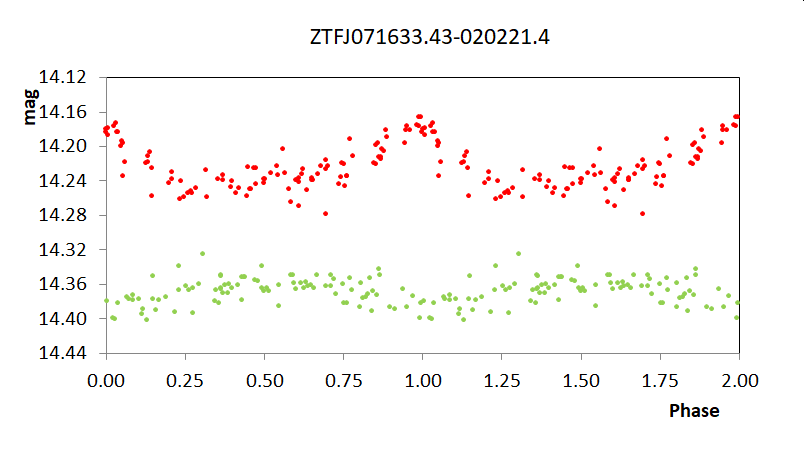}} &
    \subfigure{\includegraphics[width=0.31\textwidth]{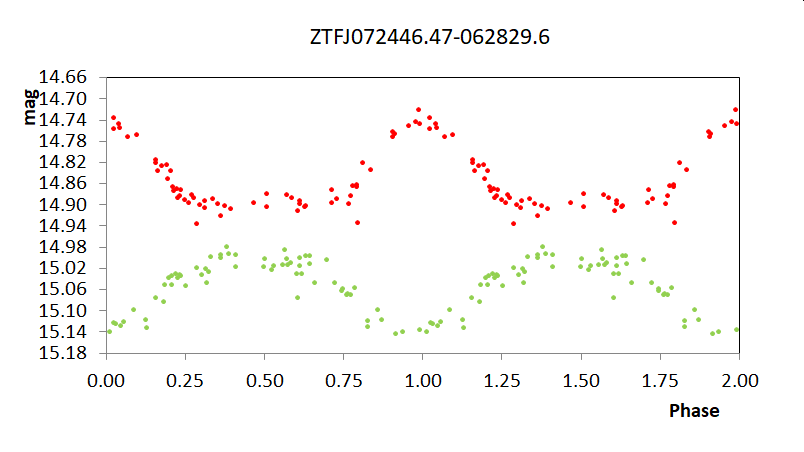}} \\
    \subfigure{\includegraphics[width=0.31\textwidth]{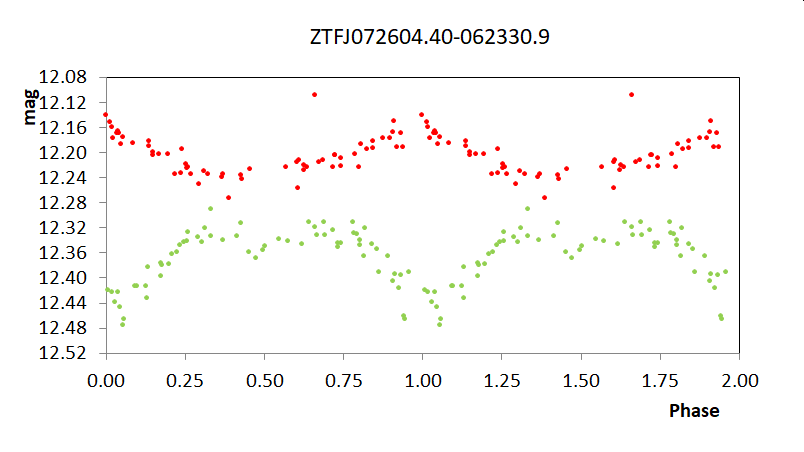}} &
    \subfigure{\includegraphics[width=0.31\textwidth]{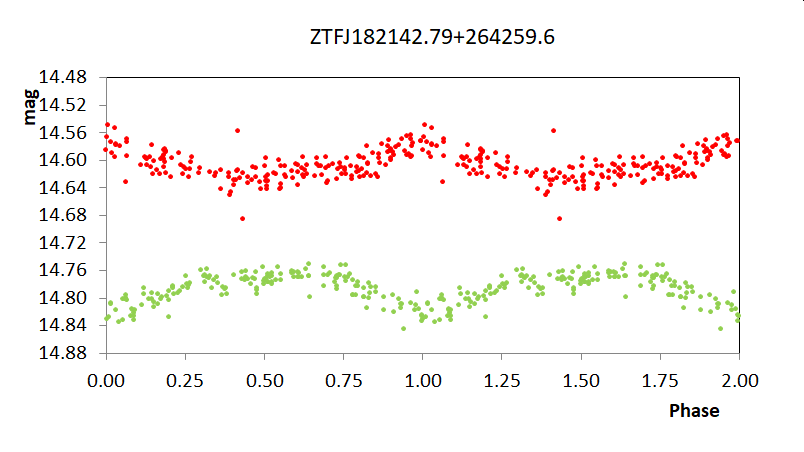}} &
    \subfigure{\includegraphics[width=0.31\textwidth]{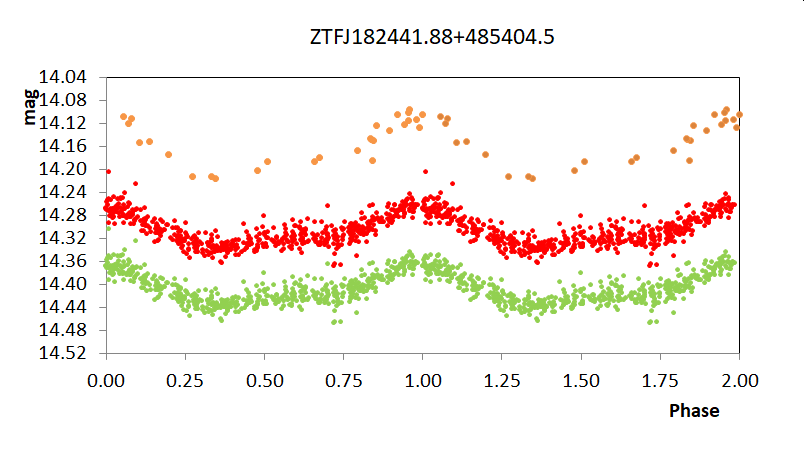}} \\
    \subfigure{\includegraphics[width=0.31\textwidth]{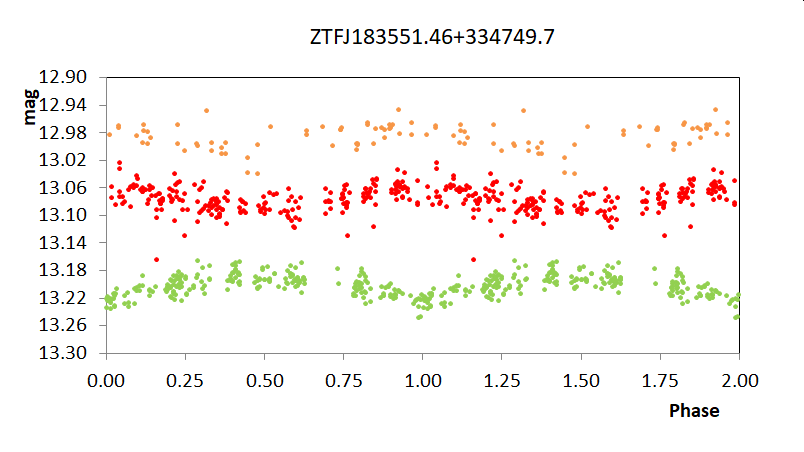}} &
    \subfigure{\includegraphics[width=0.31\textwidth]{41.png}} &
    \subfigure{\includegraphics[width=0.31\textwidth]{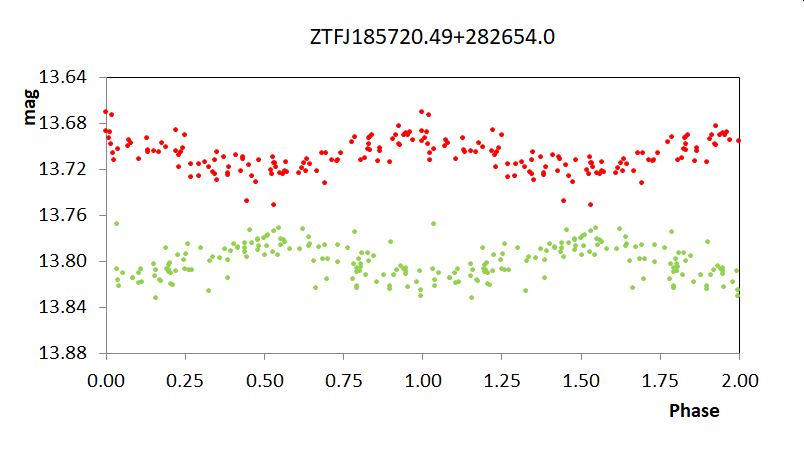}} \\
	 \end{tabular}
   \caption{continued.}
\end{figure*}
\addtocounter{figure}{-1}
\begin{figure*}
   \centering
	 \begin{tabular}{ccc}
	\subfigure{\includegraphics[width=0.31\textwidth]{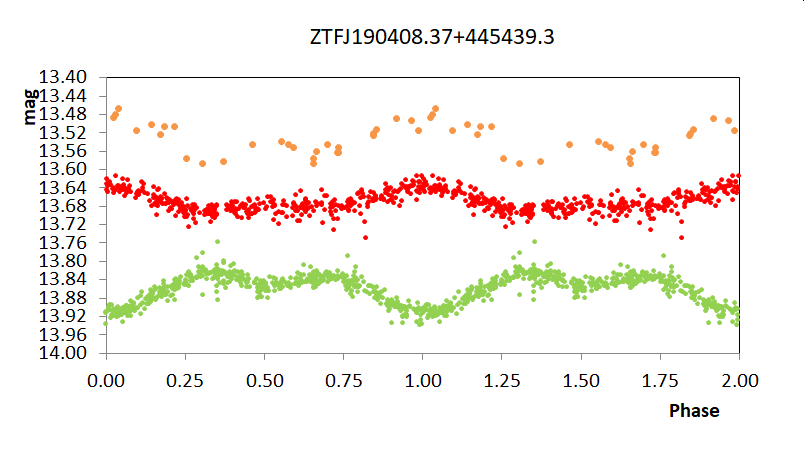}} &
    \subfigure{\includegraphics[width=0.31\textwidth]{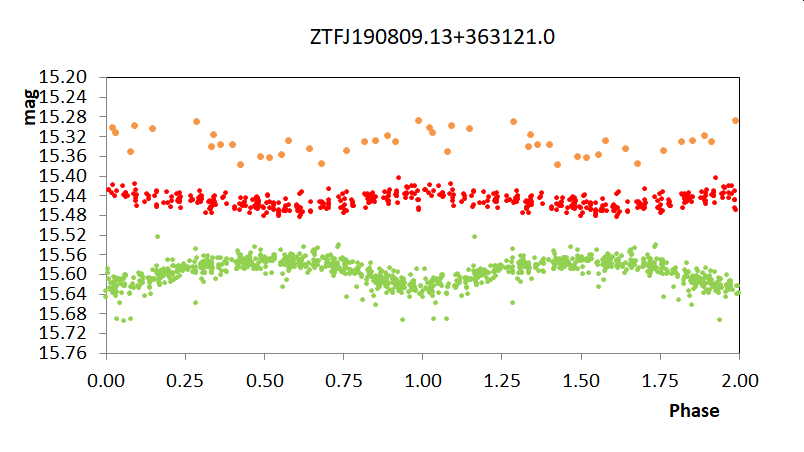}} &
    \subfigure{\includegraphics[width=0.31\textwidth]{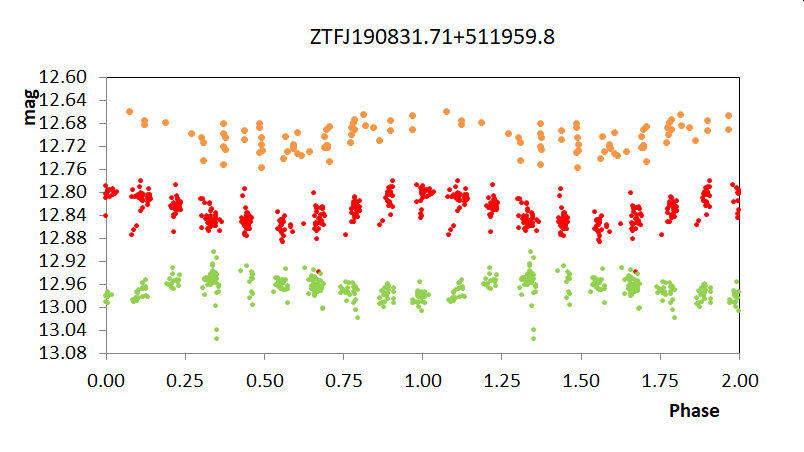}} \\
    \subfigure{\includegraphics[width=0.31\textwidth]{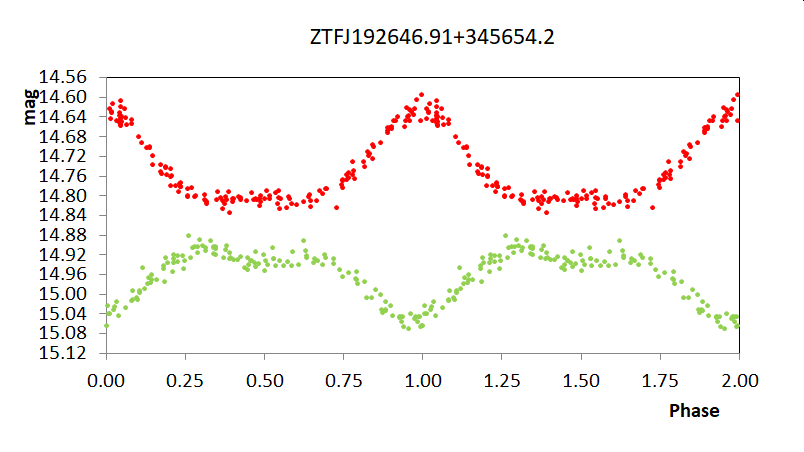}} &
    \subfigure{\includegraphics[width=0.31\textwidth]{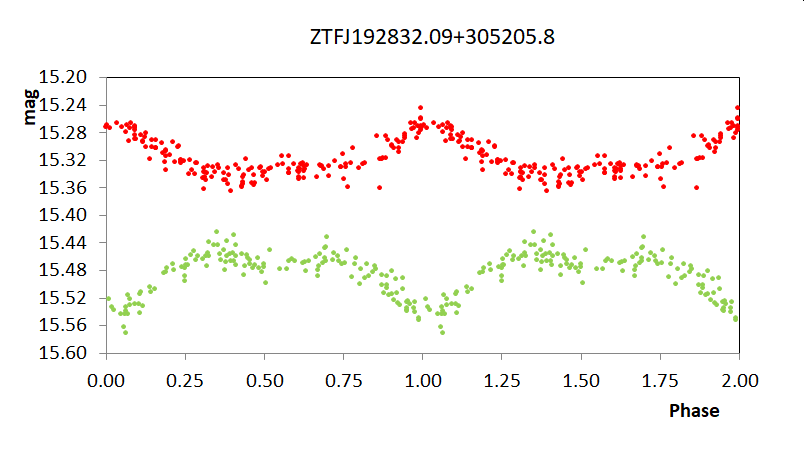}} &
    \subfigure{\includegraphics[width=0.31\textwidth]{48.png}} \\
    \subfigure{\includegraphics[width=0.31\textwidth]{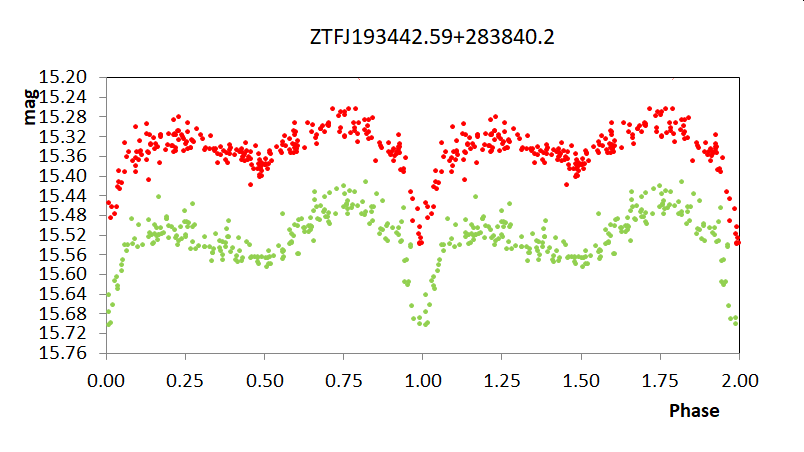}} &
    \subfigure{\includegraphics[width=0.31\textwidth]{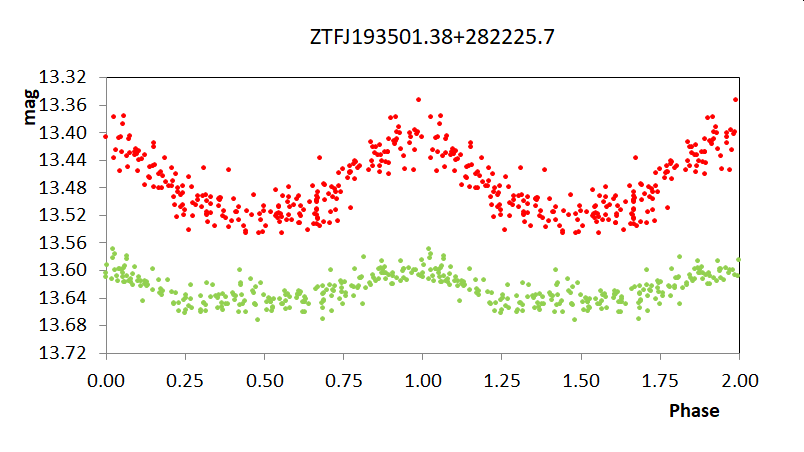}} &
    \subfigure{\includegraphics[width=0.31\textwidth]{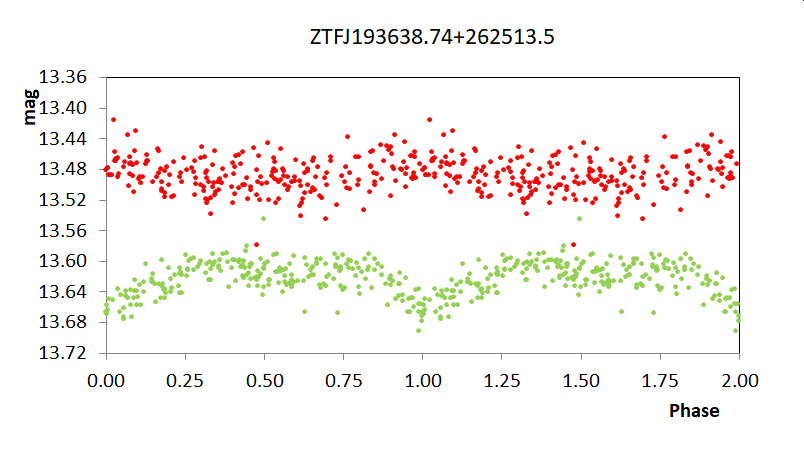}} \\
    \subfigure{\includegraphics[width=0.31\textwidth]{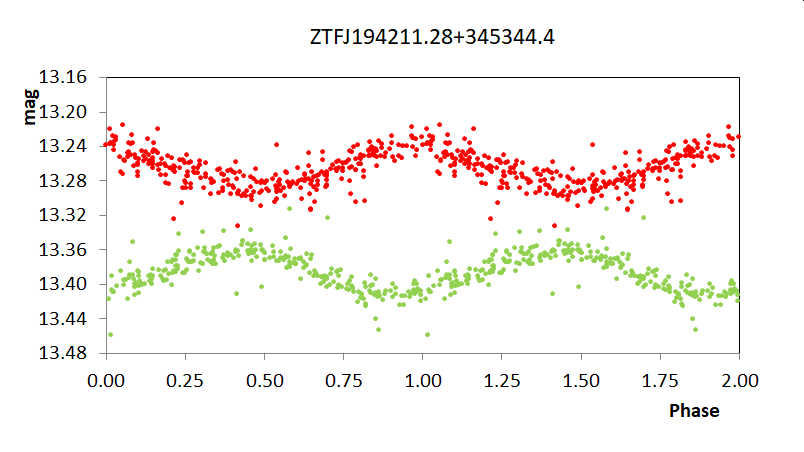}} &
    \subfigure{\includegraphics[width=0.31\textwidth]{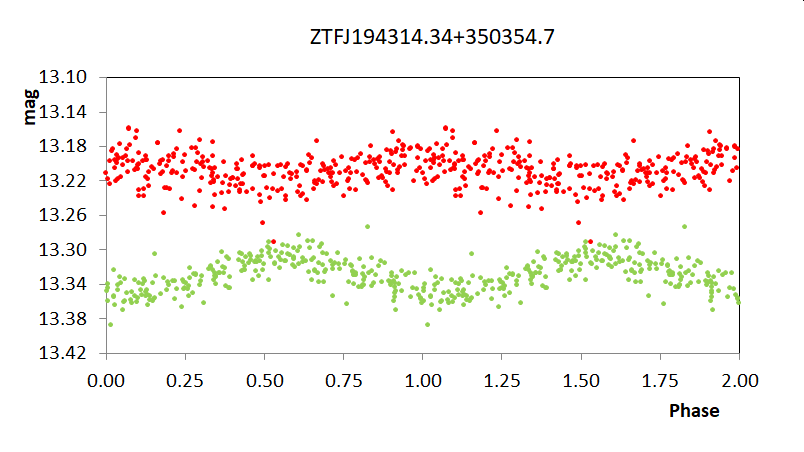}} &
    \subfigure{\includegraphics[width=0.31\textwidth]{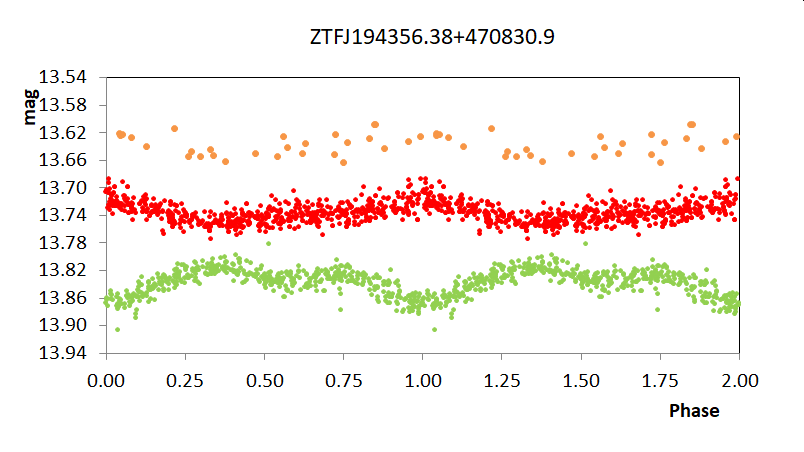}} \\
    \subfigure{\includegraphics[width=0.31\textwidth]{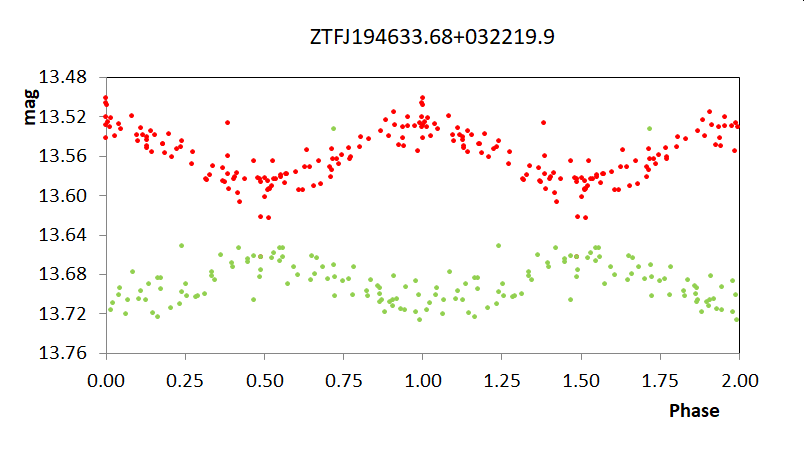}} &
    \subfigure{\includegraphics[width=0.31\textwidth]{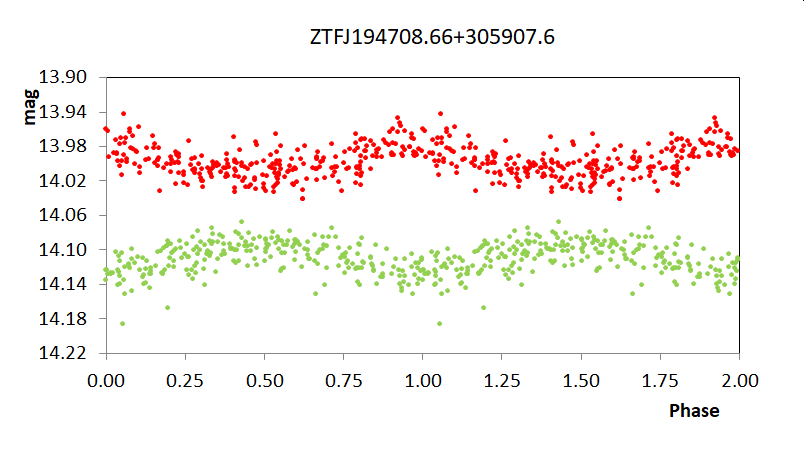}} &
    \subfigure{\includegraphics[width=0.31\textwidth]{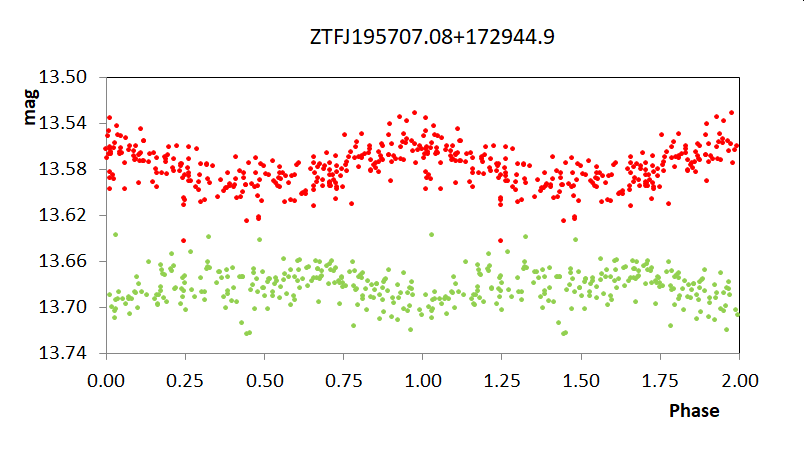}} \\
    \subfigure{\includegraphics[width=0.31\textwidth]{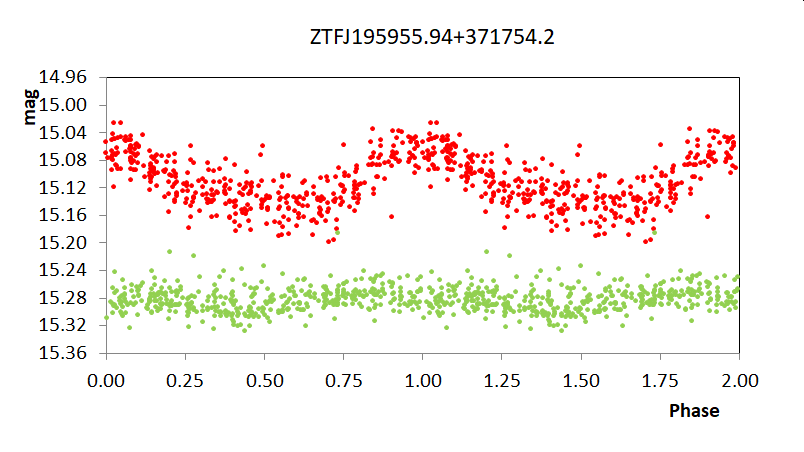}} &
    \subfigure{\includegraphics[width=0.31\textwidth]{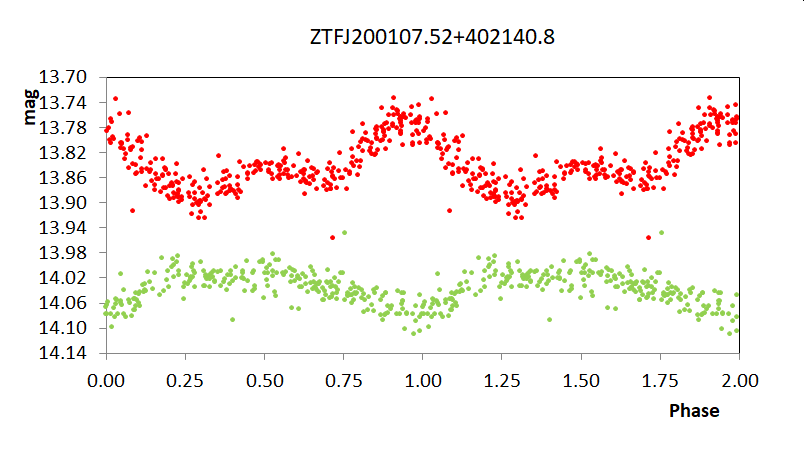}} &
    \subfigure{\includegraphics[width=0.31\textwidth]{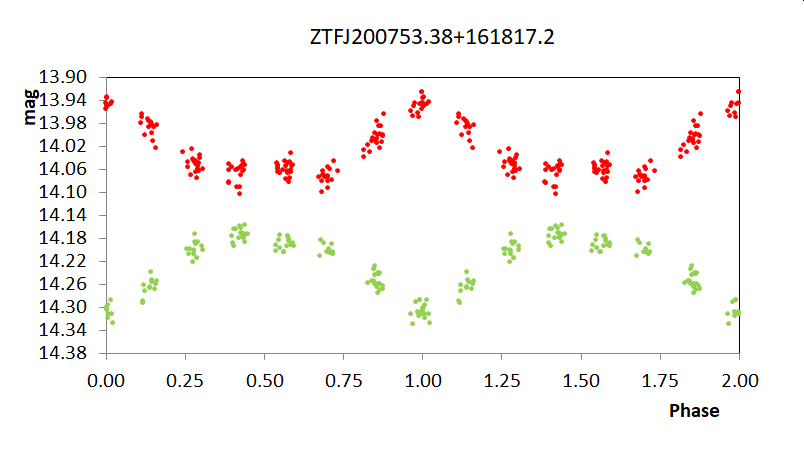}} \\
    \subfigure{\includegraphics[width=0.31\textwidth]{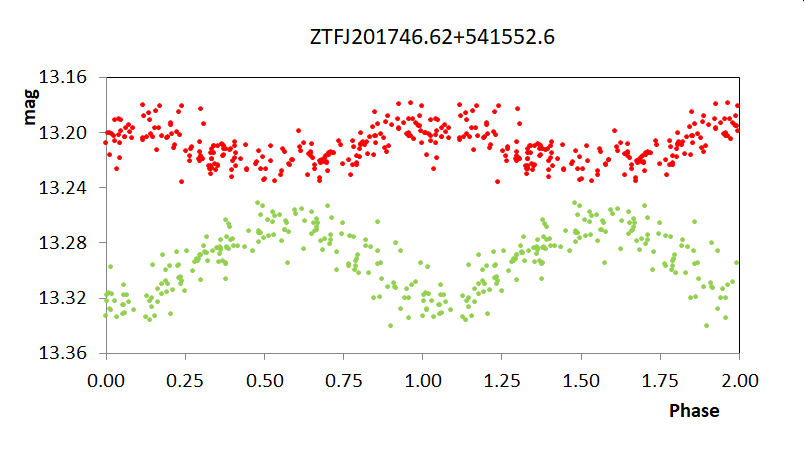}} &
    \subfigure{\includegraphics[width=0.31\textwidth]{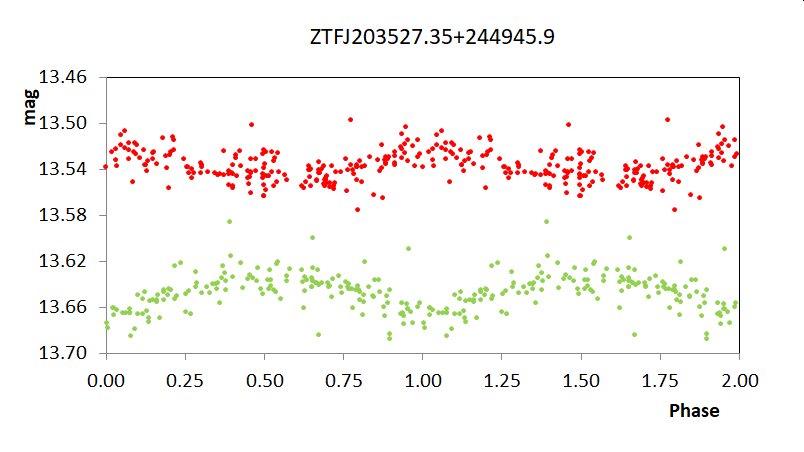}} &
    \subfigure{\includegraphics[width=0.31\textwidth]{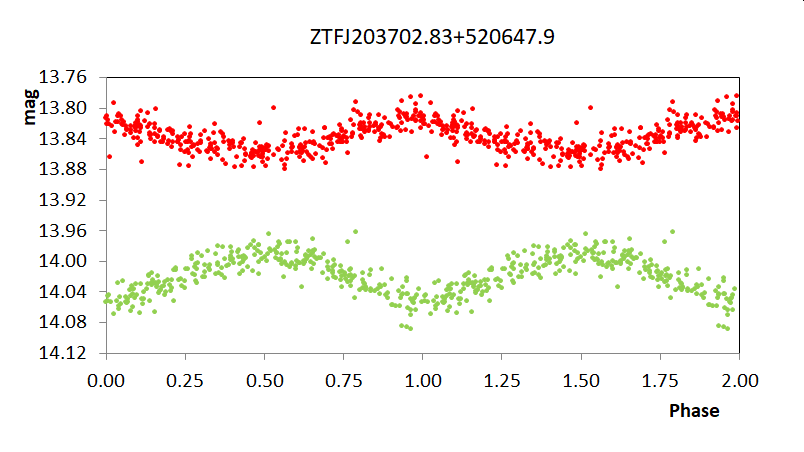}} \\
	 \end{tabular}
   \caption{continued.}
\end{figure*}
\addtocounter{figure}{-1}
\begin{figure*}
   \centering
	 \begin{tabular}{ccc}
	\subfigure{\includegraphics[width=0.31\textwidth]{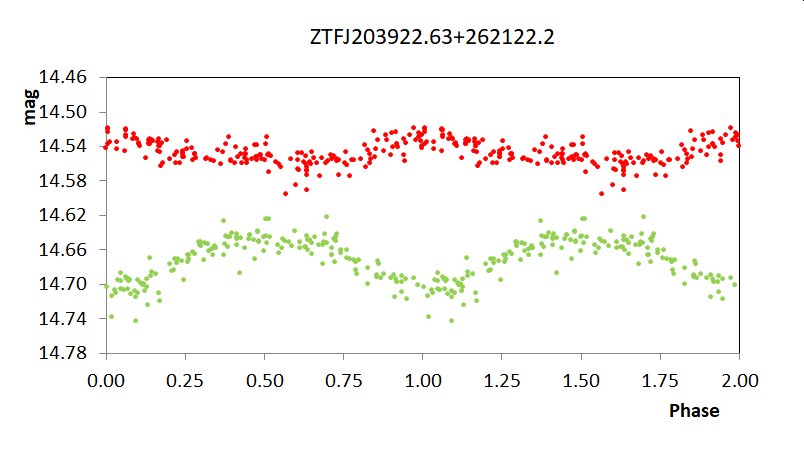}} &
    \subfigure{\includegraphics[width=0.31\textwidth]{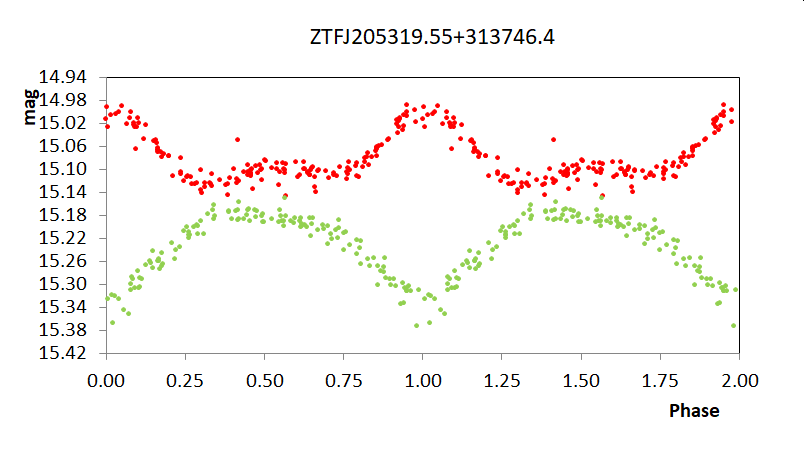}} &
    \subfigure{\includegraphics[width=0.31\textwidth]{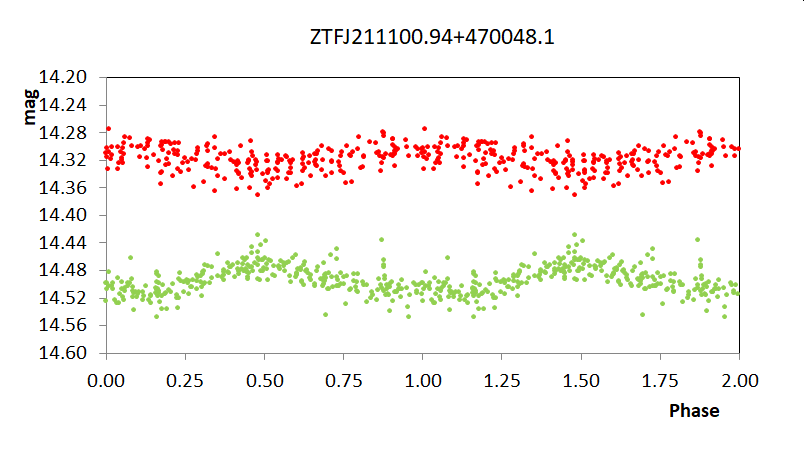}} \\
    \subfigure{\includegraphics[width=0.31\textwidth]{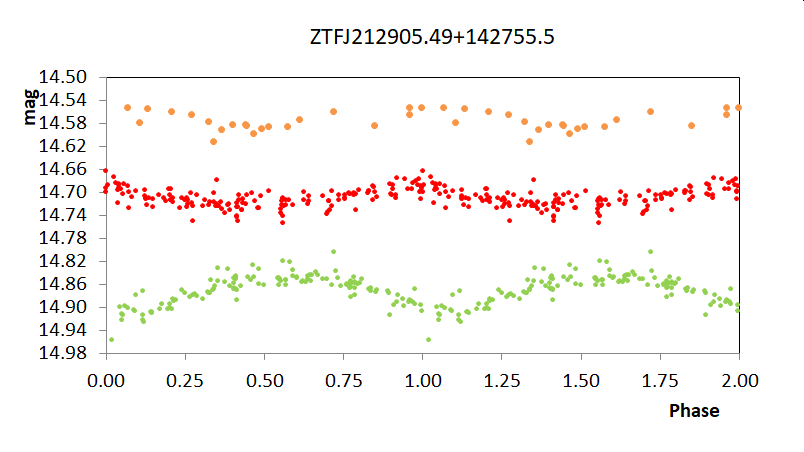}} &
    \subfigure{\includegraphics[width=0.31\textwidth]{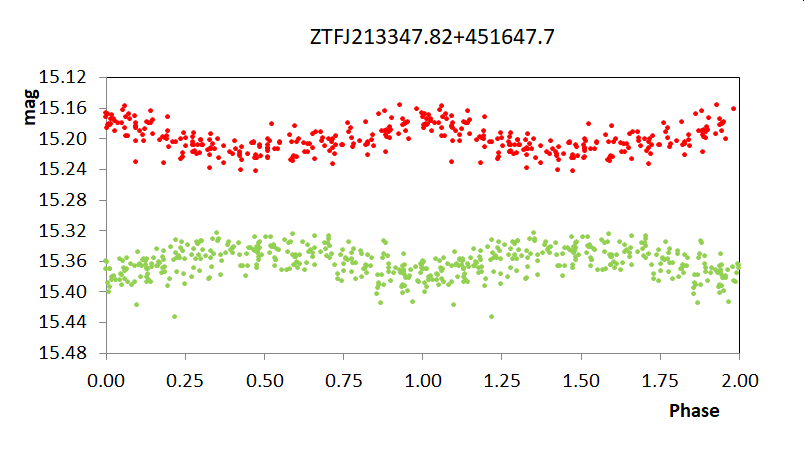}} &
    \subfigure{\includegraphics[width=0.31\textwidth]{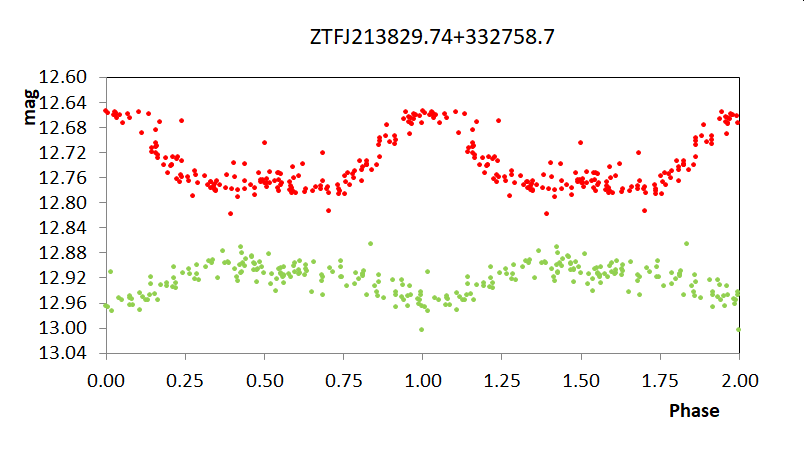}} \\
    \subfigure{\includegraphics[width=0.31\textwidth]{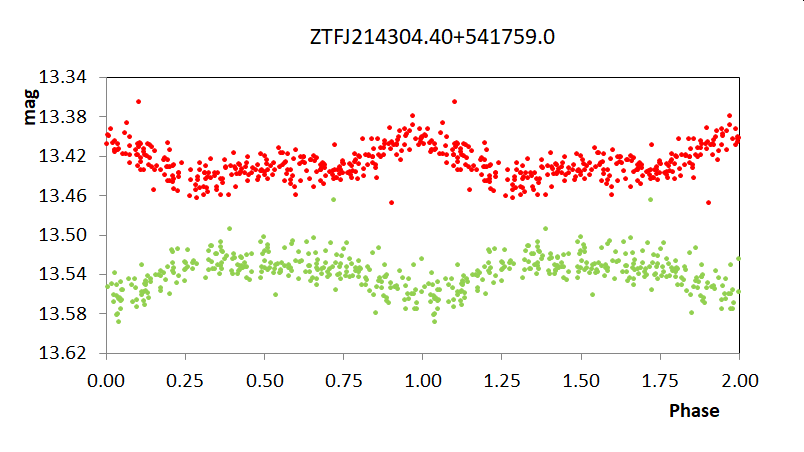}} &
    \subfigure{\includegraphics[width=0.31\textwidth]{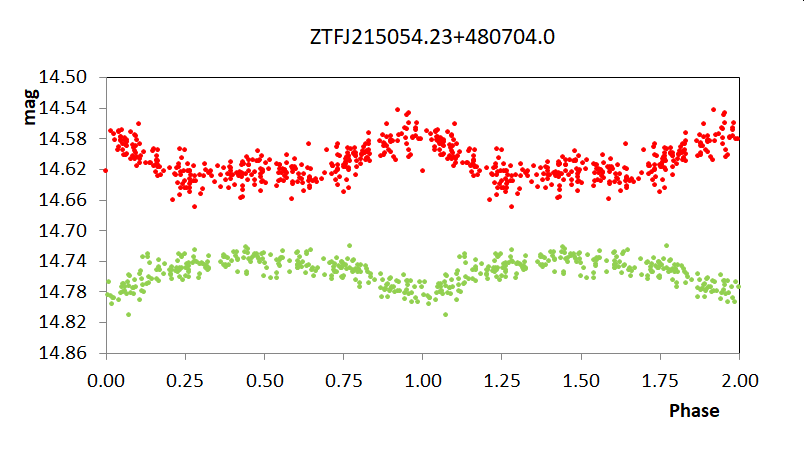}} &
    \subfigure{\includegraphics[width=0.31\textwidth]{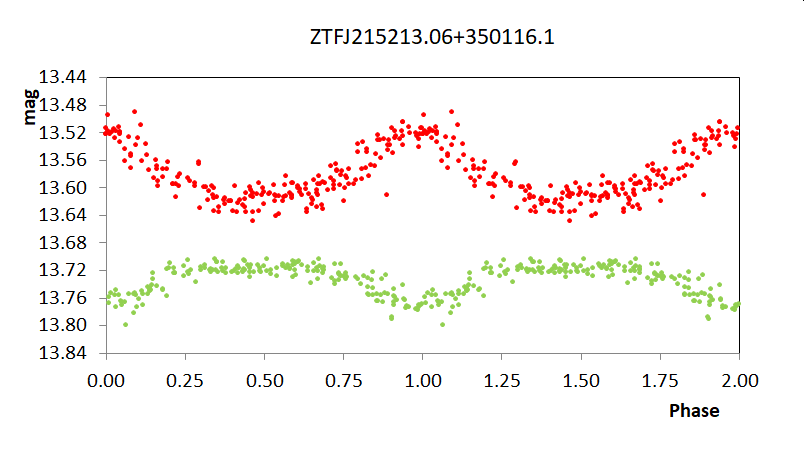}} \\
    \subfigure{\includegraphics[width=0.31\textwidth]{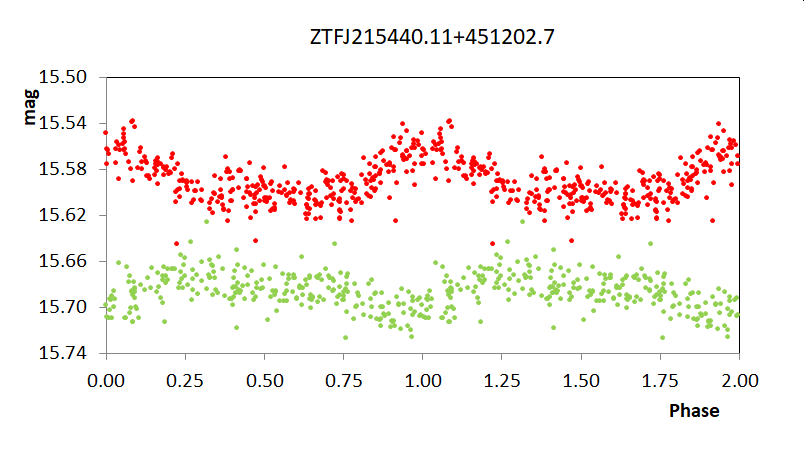}} &
    \subfigure{\includegraphics[width=0.31\textwidth]{74.png}} &
    \subfigure{\includegraphics[width=0.31\textwidth]{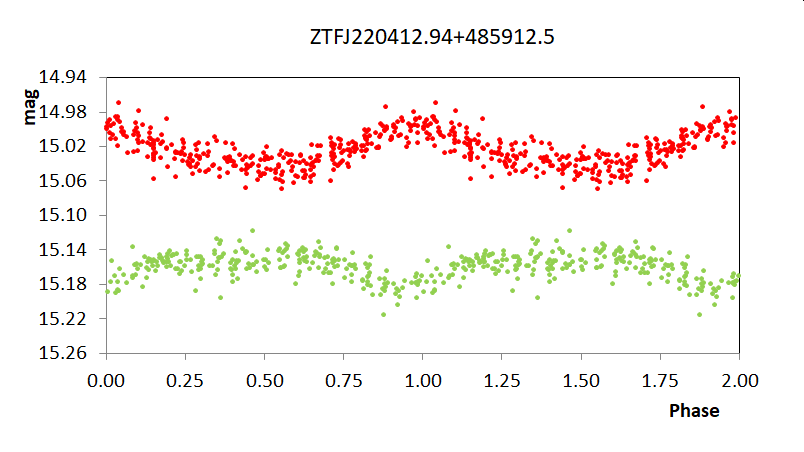}} \\
    \subfigure{\includegraphics[width=0.31\textwidth]{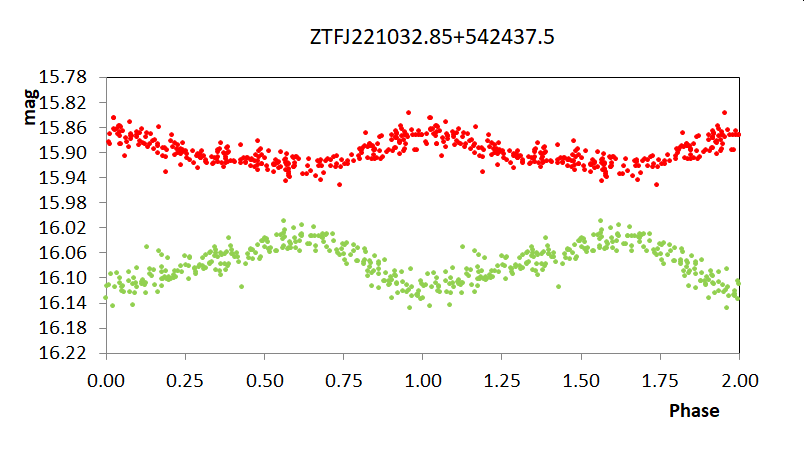}} &
    \subfigure{\includegraphics[width=0.31\textwidth]{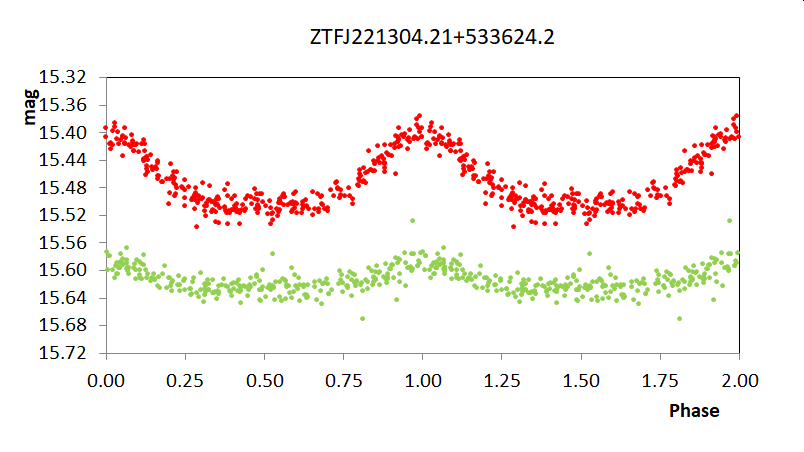}} &
    \subfigure{\includegraphics[width=0.31\textwidth]{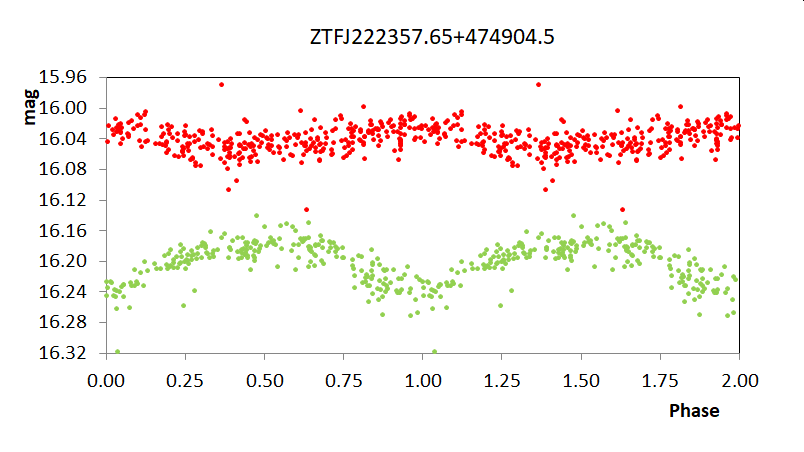}} \\
    \subfigure{\includegraphics[width=0.31\textwidth]{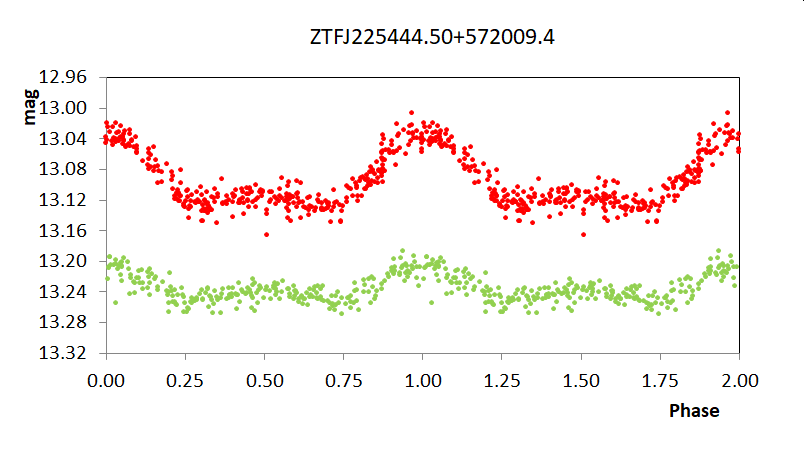}} &
    \subfigure{\includegraphics[width=0.31\textwidth]{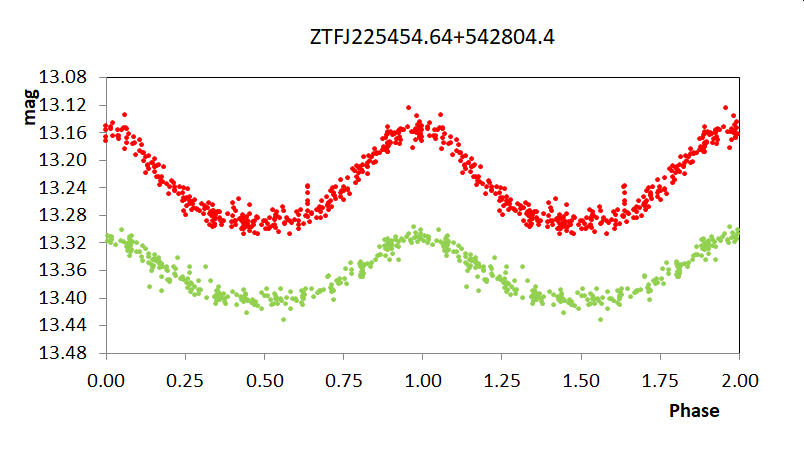}} &
    \subfigure{\includegraphics[width=0.31\textwidth]{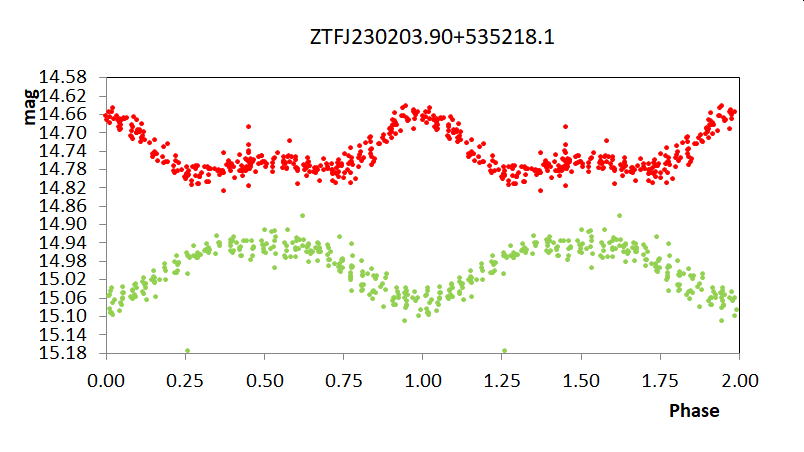}} \\
    \subfigure{\includegraphics[width=0.31\textwidth]{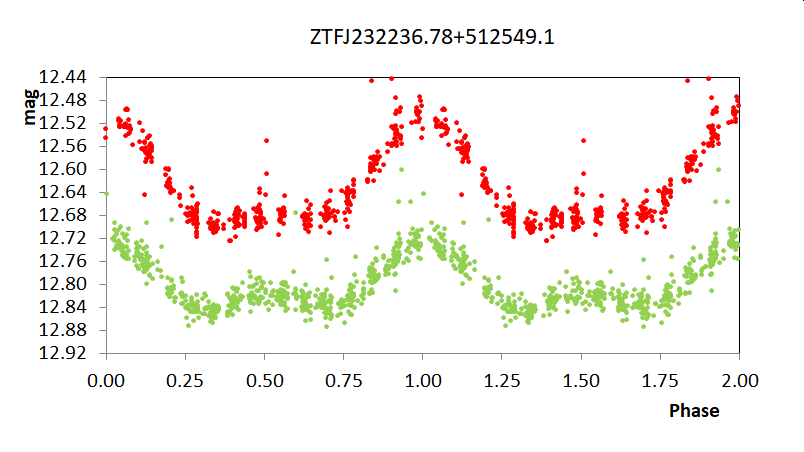}} &
    \subfigure{\includegraphics[width=0.31\textwidth]{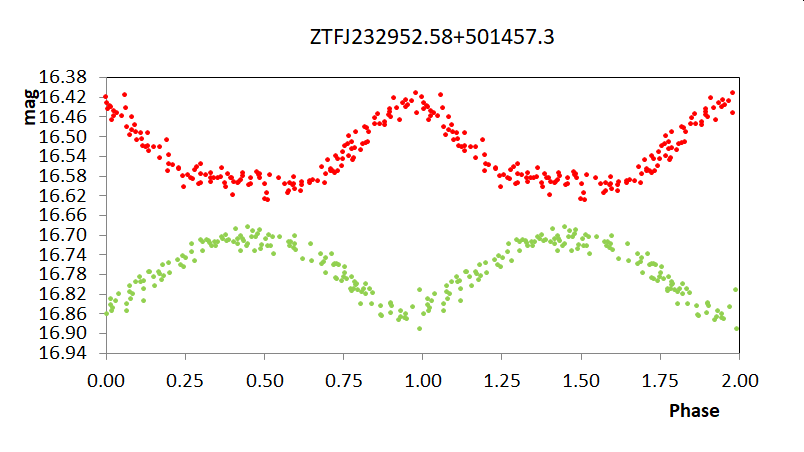}} &
    \subfigure{\includegraphics[width=0.31\textwidth]{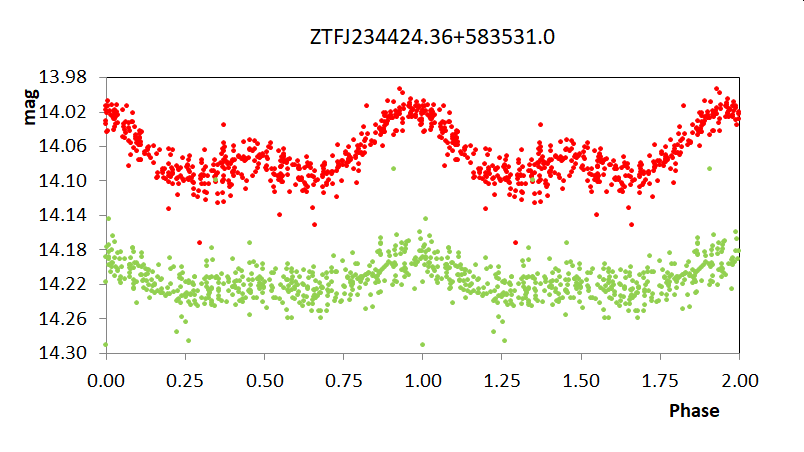}} \\
	 \end{tabular}
   \caption{continued.}
\end{figure*}
\addtocounter{figure}{-1}
\begin{figure*}
   \centering
	 \begin{tabular}{ccc}
	\subfigure{\includegraphics[width=0.31\textwidth]{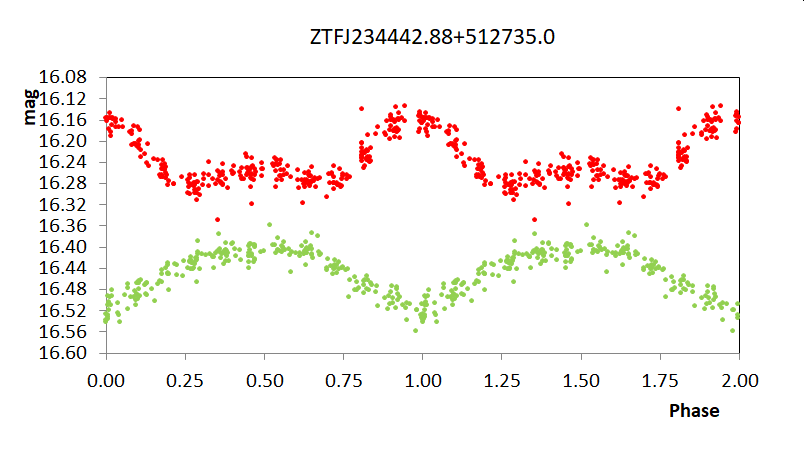}} &
    \subfigure{\includegraphics[width=0.31\textwidth]{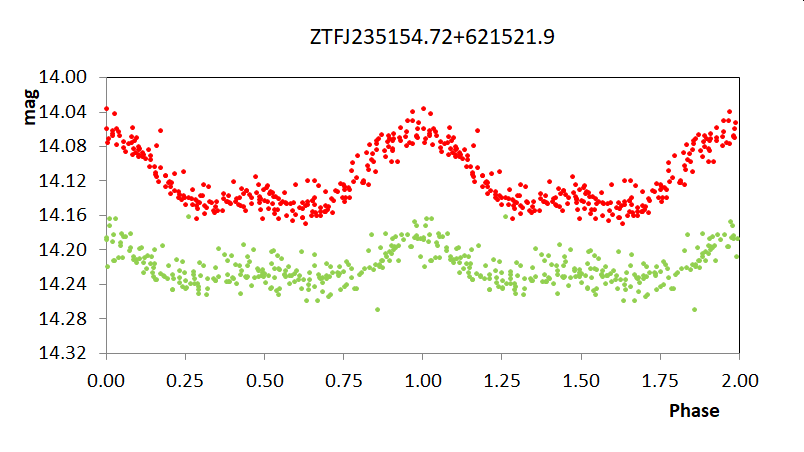}} \\
	 \end{tabular}
   \caption{continued.}
\end{figure*}

\end{appendix}

\end{document}